%% file: paper.tex
\documentclass[letterpaper, conference]{IEEEtran}
\usepackage{ulem}

\usepackage{amsmath}
\usepackage{epsfig}
\usepackage{wrapfig}
\usepackage{multicol}
\usepackage{soul}
\usepackage{multirow}
\usepackage{color}
\usepackage{psfrag}
\usepackage[linesnumbered,ruled,commentsnumbered]{algorithm2e}
\usepackage{subfigure}
\usepackage{amsmath}
\usepackage{algpseudocode}

\usepackage{tabularx,booktabs}
\newcolumntype{Y}{>{\centering\arraybackslash}X}
\usepackage{multirow}
\usepackage[table,xcdraw]{xcolor}
\usepackage{url}
\usepackage{hyperref}
\usepackage{path}
\usepackage{caption}
\usepackage{adjustbox}
\usepackage{courier}
\usepackage{listings}
\usepackage{pifont}
\usepackage{graphicx}
\usepackage{comment} 
\usepackage{epstopdf}
\DeclareGraphicsExtensions{.eps}

\usepackage{float}
\usepackage{lipsum}
\usepackage{hyperref}

\usepackage{textcomp}
\usepackage{titlecaps}
\usepackage{setspace}
\usepackage{tabularx}
\newcolumntype{P}[1]{>{\centering\arraybackslash}p{#1}}
\newcolumntype{M}[1]{>{\centering\arraybackslash}m{#1}}

\newcommand{\squishlist}{
  \begin{list}{$\bullet$}
  { \setlength{\itemsep}{0pt}      \setlength{\parsep}{-0pt}
    \setlength{\topsep}{4pt}       \setlength{\partopsep}{0pt}
    \setlength{\listparindent}{-2pt}
    \setlength{\itemindent}{-5pt}
    \setlength{\leftmargin}{1em} \setlength{\labelwidth}{0em}
    \setlength{\labelsep}{0.5em} } }

\newcommand{\squishend}{
    \end{list}  }

\newcommand{\cb}{\textcolor{blue}}
\newcommand{\cred}{\textcolor{red}}
\newcommand{\cg}{\textcolor{green}}

\let\oldnl\nl% Store \nl in \oldnl
\newcommand{\nonl}{\renewcommand{\nl}{\let\nl\oldnl}}% Remove line number for one line
\newcommand{\tbstore}{\textit{\titlecap{\scshape SciSpace}\normalfont}}

\usepackage{ulem}

\ifCLASSINFOpdf
\else
\fi
\begin{document}
\title{\huge 
  \tbstore{}: A Scientific Collaboration Workspace for File Systems in Geo-Distributed HPC Data Centers }

%\author{\IEEEauthorblockN{Michael Shell}
%\IEEEauthorblockA{School of Electrical and\\Computer Engineering\\
%Georgia Institute of Technology\\
%Atlanta, Georgia 30332--0250\\
%Email: http://www.michaelshell.org/contact.html}
%\and
%\IEEEauthorblockN{Homer Simpson}
%\IEEEauthorblockA{Twentieth Century Fox\\
%\and
%\IEEEauthorblockN{James Kirk\\ and Montgomery Scott}
%\IEEEauthorblockA{Starfleet Academy\\
%San Francisco, California 96678--2391\\

\author{
	Awais Khan$^{1}$, Taeuk Kim$^{1}$, Hyunki Byun$^{1}$, Youngjae Kim$^{1}$%$^{\ast}$\thanks{$^{\ast}$Corresponding Author.}
	, Sungyong Park$^{1}$, Hyogi Sim$^{2}$\\
%	$^{1}$Department of Computer Science and Engineering, Sogang University, Seoul, Republic of Korea\\
%	 $^{2}$Department of Computer Engineering, Ajou University, Suwon, Republic of Korea\\
	$^{1}$Sogang University, Seoul, South Korea, $^{2}$Oak Ridge National Laboratory, Oak Ridge, TN, USA\\

         {\small\{awais, taeugi323, bhyunki, youkim, parksy\}@sogang.ac.kr}, \small{simh}@ornl.gov      		
}


%Telephone: (800) 555--1212, Fax: (888) 555--1212}


\maketitle

\input{abstract}

%\saythanks

% \fi
\IEEEpeerreviewmaketitle

\setstretch{0.95}

\vspace{-0.05in}
\input{intro}

\vspace{-0.05in}
\input{motiv}
\vspace{-0.05in}
\input{overview}

\vspace{-0.05in}
\input{design}
\vspace{-0.05in}
\input{eval}

\vspace{-0.05in}
\input{conc}
\vspace{-0.05in}

\let\OLDthebibliography\thebibliography
\renewcommand\thebibliography[1]{
  \OLDthebibliography{#1}
  \setlength{\parskip}{0pt}
  \setlength{\itemsep}{0pt}
 %\footnotesize
 %\tiny
 \scriptsize
}

%\tiny

\normalem

\bibliographystyle{IEEETran}
\setstretch{0.84}
\bibliography{refs}

\end{document}

%% file: abstract.tex
\begin{abstract}
	
%The scientific communities are accelerating the trend to constitute a \textit{Small World}, the collaboration among geo-distributed data centers to enhance the information and resource sharing for joint simulation and analysis.  Such data centers allow access to their storage and compute clusters to collaborators via multiple data transfer nodes (DTNs). However, this approach includes no automation and is quite inconvenient to collaborators for viewing and accessing the data through multiple DTNs, as a single unified namespace does not exist. Moreover, the existing data transfer tools limits the scientific collaborations due to manual data transfer management.

Future terabit networks are committed to dramatically improving big data motion between geographically dispersed HPC data centers. The scientific community takes advantage of the terabit networks such as DOE's ESnet and accelerates
%the scientific community is accelerating 
the trend to build a small world of collaboration between geospatial HPC data centers. 
%to 
%\cred{\sout{It eventually improves 
%improving 
It improves information and resource sharing for joint simulation and analysis between the HPC data centers.
In this paper, we propose 
%an idea 
to build \textit{\tbstore{}} (Scientific Collaboration Workspace) for collaborative data centers. It provides a global view of information shared from multiple geo-distributed HPC data centers under a single workspace. 
%\tbstore{} offers a transparent POSIX-compatible namespace and it can be adopted by existing scientific applications and file systems without modifications. 
\tbstore{} supports native data-access to gain high-performance when data read or write is required in native data center namespace. 
It is accomplished by integrating 
%an extensive 
a metadata export protocol. 
To optimize scientific collaborations across HPC data centers, 
%we integrate 
\tbstore{} implements search and discovery service. 
%search and discovery service is integrated in \tbstore{}. 
%offering
%and offer 
%multiple data extraction modes based on performance metrics. 
To evaluate, we configured two geo-distributed small-scale HPC data centers connected via high-speed Infiniband network, equipped with LustreFS. 
We show the feasibility of \tbstore{} using real scientific datasets and applications. 
The evaluation results show average 36\% performance boost when the proposed native-data access is employed in collaborations.

\end{abstract}

%% file: intro.tex
\section{Introduction}
\label{sec:intro}
\vspace{-0.05in}
In recent years, we are experiencing a data explosion: almost 90\% of today\textquotesingle s data has been produced in the last two years, with data being produced in the magnitude of petabytes~\cite{thusoo2010hive}. A weather company reported that more than 20 terabytes of data being generated each day for storing temperature readings, wind speeds, barometric pressures and satellite images across the globe~\cite{scidb}. Several DOE's HPC (High Performance Computing) leadership-computing 
facilities, such as 
%% hs: explicitly write names when first time mentioned
OLCF (Oak Ridge Leadership Computing Facility), 
NERSC (National Energy Research Scientific Computing Center), 
and ALCF (Argonne Leadership Computing Facility),
%generate hundreds of petabytes per year of simulation data 
generate hundreds of petabytes of simulation data annually
and are projected to generate in excess of
one
exabyte per year~\cite{extreme}. 
To accommodate such growing volumes of data, 
science and %scientific and 
research communities are deploying larger, well-provisioned geo-distributed storage and computation HPC clusters~\cite{extreme}.

%Such clusters are generally deployed as separate sets of storage and compute nodes, providing massive numbers of CPU cores, low latency interconnects, as well as fast concurrent data bandwidth typically managed by an underlying parallel file system~\cite{extreme, IOChallenge, Schmuck:2002:GSF:1083323.1083349, Shvachko:2010:HDF:1913798.1914427}. 

%Figure~\ref{fig:motivation} shows the high-end decoupled storage and computing architecture of two geo-distributed data centers connected via the high-speed network such as terabit network infrastructure in DOE\textquotesingle s ESnet~\cite{esnet, Kim:2015:LOD:2750482.2750488}, each data center equipped with the parallel file system such as Lustre~\cite{lustre}. 

%% hs: I do not understand this.

%However \cb{(
%Usually, 
In the HPC data centers, 
%to mitigate security risks, 
data transfer nodes (DTNs) are
supplied to access the provisioned storage and compute clusters~\cite{dtn}.
%The use of DTNs mitigates security risks. 
External access using DTN mitigates security risks.
%\cred{However, to keep architectural abstraction, data transfer nodes (DTNs) are
%supplied to access the provisioned storage and compute clusters.}
Atop %of
such DTNs, scientists and researchers across different HPC data centers collaborate by
sharing simulation and analytical data for 
%quality %\cb{Taeuk : quality of}
science 
%\cred{hs: quality also can be adjective} 
research and discovery
%services
~\cite{klimatic, cooperative}. 
%The collaborations are motivated in such
%scenarios due to high-speed terabit network availability. 
Particularly, the high-speed terabit network connections between HPC data centers expedite such collaborations.
%In specific, 
DOE\textquotesingle s
ESnet currently supports 100~Gb/s of data transfers between DOE facilities.
%and, in future deployments, %will most likely 
In future deployments, it is expected to
support 400 Gb/s followed by 1Tbps~\cite{esnet}.
%\cred{Explain that, as we mentioned in the abstract, the reason why collaboration between data centers is possible is because DOE data centers are connected to terabits networks such as ESNet! } \cb{Khan: Addressed}
%For example, 
Generally, scientists and their collaborators using the DOE facilities typically have access to additional storage and compute resources 
%\cred{Kim: what are these resources?}
at multiple geo-distributed HPC %facilities and 
data centers. %\cb{(hyunki:generally and typically looks redundant)}
%They utilize those resources to simulate and analyze data generated
%from experiment facilities on supercomputers and validate results.
By exploiting various computing resources at geo-dispersed HPC data centers, scientists 
%can 
efficiently
perform simulations and data analyses, resulting in fast scientific discoveries.
%Other models of large collaborations include: 
For instance, an OLCF petascale simulation needs
nuclear interaction datasets processed at
NERSC~\cite{Kim:2015:LOD:2750482.2750488}.
%the ALCF runs a climate simulation
%and validates the simulation results with climate observation datasets at ORNL
%data center~\cite{Kim:2015:LOD:2750482.2750488}. 
Similarly, scientists in ALCF validate their simulation results by comparing them with climate observation datasets at ORNL data center~\cite{Kim:2015:LOD:2750482.2750488}. 
This collaboration between data centers is accompanied by data movement between OLCF and ALCF.

A traditional %scientific workflows adopted in collaborations 
workflow of scientific collaborations %are 
is as follows:
the scientists at different facilities engage remote access tools such as SSH
to connect remote sites and find the required datasets, copy the datasets to
local sites via data transfer tools such as bbcp and scp, and afterwards,
execute the analysis~\cite{Kim:2015:LOD:2750482.2750488}.
Figure~\ref{fig:motivation} depicts %the likewise 
the traditional
collaboration model %activity 
between
two collaborators %at 
from 
%two 
different
%geo-distributed 
HPC data centers. However, such an approach %suffers %when the data centers in collaboration are more than one. 
does not work when multiple HPC data centers are involved, because
%Collaborators cannot
%view the data shared from multiple sites in a single workspace via SSH.}
a SSH session is unable to present a single, unified workspace out of
all shared datasets from multiple data centers.
%So 
Therefore,
it is %highly required 
crucial to render a unified %common 
view of shared datasets to all the
collaborators via a collaborative namespace. 
%The 
Existing studies, such as CFS~\cite{cooperative}, OceanStore~\cite{oceanstore}, Campaign Storage~\cite{campaign}, and UnionFS~\cite{unionfs} %are centered on granting 
can provide
an aggregated storage space, but %and 
not the collaborative namespace. 
%The lack of the collaborative namespace forces scientists to
%manually copy and publish datasets within the global aggregated namespace,
%precluding an ability to configure sharing status of individual datasets when required.
%\cb{(hyunki:What is difference between aggregated storage namespace and collaborative space? I also think it's little ambiguous.)}
%In particular, 
The collaborative namespace in \tbstore{} eliminates
the need for laborious data transfers and managements, which have been conducted manually by scientists,
by allowing fine-grained sharing configurations for individual datasets.

Above all, scientists in collaboration might require analyzing the specific
datasets based on certain conditions, 
for example, \textit{an analysis on a dataset %the data 
which is generated from %the 
a satellite at a certain location %ranging from
for a specific period of time, e.g., from a start point to an end point. }
%start period to the end period}. 
%In general, the file systems itself 
%do not offer any kind of such a query-cooperative search feature. 
Existing parallel and distributed file systems do not directly support such advanced, data-aware search queries.
A common approach to
%adopt 
provide the advanced data
search service is to build a metadata indexing %an index management 
layer, using an external database system, %(external database)
%placed 
between the application and the file system. 
%Whereas, it requires modifications at both application and file system level to
%integrate search feature and requires scientists to learn SQL APIs (Non-POSIX).
However, this not only requires modifications to both applications and the file system, but also forces scientists to use the SQL interface instead of the familiar file system interface.
TagIT~\cite{tagit} offers data extraction and discovery service on top of a file
system namespace. However, the solution is heavily dependent on the GlusterFS architecture
and unable to index legacy datasets.
%However, limitation includes tight coupling with GlustreFS
%architecture and do not consider data indexing on legacy data. 
%\cred{Introduce
%only the TagIt paper and explain its disadvantages and explain how we did it.
%Klimatic and VSFS will be introduced in the related work.} %\cb{Khan: Adddressed}
%Klimatic~\cite{klimatic}, TagIT~\cite{tagit} and VSFS~\cite{vsfs} offer data extraction and discovery service on top of general file system namespace. The VSFS offers DRAM-based distributed architecture inheriting the DRAM size limitation and is not failure-tolerant~\cite{vsfs}. 
%Both approaches  
Likewise, a single scientist %collaborator 
can be involved in multiple, separate or
overlapping collaborations, which is not addressed by any of the existing
studies. 
%Thus, we claim that %it is the essential time that an unusual
%collaboration workspace model be proposed 
Therefore, it is essential to provide a collaboration workspace model
to allow
%for 
practical and powerful collaborations 
in 
a paradigm where HPC data centers are connected to high-speed networks.

To address the %above-mentioned 
aforementioned
challenges, we propose to build \tbstore{}, a
scientific collaboration workspace framework for file systems across
%geo-located 
geo-distributed
HPC data centers connected via the high-speed network. 
%
%\tbstore{} is envisioned to have the following contributions:
%This paper has the following contributions: 
Specifically, this paper makes the following contributions:
% contribution includes
\squishlist
\item
\tbstore{} promotes the collaboration activities among the scientists at remote
HPC sites for data sharing, joint simulation, and analysis. 
The proposed service
framework provides a virtual abstraction on top of multiple dissimilar file
systems and presents a global unified view of shared datasets %picture of shared data 
to all the collaborators.
%To minimize existing application and file system changes and to gain
%high-performance, we allow local writes and native data access. 
\tbstore{} allows a native data access %\cb{(hyunki: using local file system)}
, e.g., local file write, which allows high performance file operations and minimizes modifications to the existing applications and file systems. 
Any changes to the local data center file system are transparently applied to the collaboration workspace by Metadata Export Utility (MEU).
%\cb{(hyunki:Metadata Export Utility can apply changes to the local file system to collaboration workspace transparently.)}
%\tbstore{} offers
%{Metadata Export Utility} (MEU) %, which synchronize local modifications to the collaboration workspace.
%the metadata of unsynced files stored via local namespace to collaboration workspace. % to
%enable data sharing. 

%\begin{itemize}
\item 
\tbstore{} offers efficient Scientific Discovery Service (SDS) integrated on
top of the collaboration workspace to facilitate the %optimize 
scientific workflow. 
Specifically,
\tbstore{}
provides a multi-mode metadata extraction service based on application's requirement. % data access patterns. 
Additionally, to allow a single scientist %single collaborator 
to participate in multiple
collaborations, \tbstore{} supports %offers 
a template namespace. 
Using the template namespace, scientists can associate data sharing options,
such as shared and private namespaces,
to individual collaborations.

\item 
We conduct a comprehensive evaluation of \tbstore{} %for our proposed ideas 
by building a
collaboration between two small-scale geo-distributed HPC data centers %equipped with
with Lustre file systems~\cite{lustre}. We compare the performance of our framework with
UnionFS~\cite{unionfs}. 
%\cb{Taeuk : Introduced existing works as CFS, OceanStore, Campaign Storage before, but experiment is done by comparing with UnionFS. It's good if a mention of UnionFS comes in third paragraph of I.Introduction}
%Specifically, in our evaluation, we use synthetic and
%real scientific datasets and applications.
In addition to synthetic datasets, we also use real scientific datasets and applications.
Our evaluation demonstrates that \tbstore{} outperforms the traditional approach by 36\% on average,
in real collaborations.
%We observed that our proposed idea
%shows an average 36\% performance improvement when employed in real
%collaborations.

   % To evaluate \tbstore{}, we used two geo-distributed data centers equipped with Lustre. We mount each Lustre on two different nodes exhibiting as DTNs. We use IOR~\cite{IOR} benchmark and real scientific HDF5 datasets to evaluate the \tbstore{}. The evaluation shows the \tbstore{}\textquotesingle s feasibility of being deployed in real collaborations. 
 %   \cred{Summarize some quantitative results from the experimental results. I copied the following contribution point from my LADS paper for your reference. Write as this writing format: 
   % \cb{We conduct a comprehensive evaluation for our proposed ideas using a file size distribution based on a snap shot of one of the file systems of Spider (the previous file system) at ORNL. We compare the performance of our framework with a widely used data transfer program, bbcp [12]. Specifically, in our evaluation with the real file distribution based workload, we observe that our framework yields a 4-5 times higher data transfer rate than bbcp when using eight threads on a node. Also, we find that with a small amount of SSD, LADS can improve further the data transfer rate by 37\% over a baseline without SSD buffering and far more cost-effectively than provisioning additional DRAM.}}

%\end{itemize}
\squishend

%% file: motiv.tex
\section{Related Work}
\label{sec:motiv}
\vspace{-0.05in}

%\cred{I suggest to call this section, Related Work.} \cb{khan: renamed.}

%\cred{\sout{This section presents the fundamental background and elaborates on our observations that help motivates the \tbstore{} research.}}

\begin{figure}[!t]
	\begin{center}
		\begin{tabular}{c@{}c@{}}
			\includegraphics[width=0.44\textwidth]{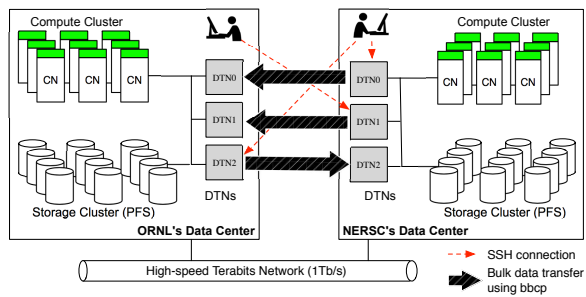}  \\
		\end{tabular}
			\vspace{-0.4cm}
		\caption{\small Scientific collaboration across two geo-distributed HPC data centers (DC) via DTNs. 
%Collaborators at different HPC data centers
%DC's 
%in collaboration via DTNs.}
}
		\label{fig:motivation}
	\end{center}
	\vspace{-0.35in}
\end{figure}

%\cred{\sout{The}} 
%\tbstore{} targets at a collaboration environment where scientists and researchers at different research and computing facilities share data to improve 
%the outcomes of scientific simulations and analytics. 
%scientific simulation and analysis results.
%In such cases, the remote collaboratos are allowed to utilize the storage and compute resources~\cite{oceanstore, campaign, cooperative}.
%\cred{Kim: Referring to references, something seems to be missing. Do not you have to mention storage?}

Figure~\ref{fig:motivation} represents a scientific collaboration environment between two geo-distributed HPC data centers, i.e., ORNL %~\cite{oakridge} 
and NERSC, %~\cite{nerc}
%are in collaboration on different projects and allow the remote collaborators
allowing the remote collaborators
to access the local %provided 
facilities~\cite{Kim:2015:LOD:2750482.2750488}.
%% hs: I don't think lustre, glusterfs can be called "studies".
%Several existing studies 
%such as GFarm~\cite{gfarmfs}, XtreemFS~\cite{xtreemfs},
%iRODS~\cite{irods}, Hadoop~\cite{Shvachko:2010:HDF:1913798.1914427},
%Ceph~\cite{weil2006ceph}, Lustre~\cite{lustre}, and
%GlustreFS~\cite{Davies:2013:SOG:2555789.2555790},
%provide an aggregate view of data stored on multiple nodes deployed on a single site, 
Existing storage systems, such as
GFarm~\cite{gfarmfs}, XtreemFS~\cite{xtreemfs},
iRODS~\cite{irods}, Hadoop~\cite{Shvachko:2010:HDF:1913798.1914427},
Ceph~\cite{weil2006ceph}, Lustre~\cite{lustre}, and
GlustreFS~\cite{Davies:2013:SOG:2555789.2555790},
can provide an aggregate view of data stored
on multiple nodes %deployed on 
within a single facility.
%however 
%such an aggregate\cb{d} storage view requires a common storage interface deployed on each node,
However,
such systems attain the aggregated storage view by deploying an identical storage interface
on each storage node 
%\cred{Kim: what is node?}
and
do not support the unification of dissimilar file systems.
%Unification of  %or gluing 
%multiple \cb{dissimilar} file systems is not considered in these studies. 
Campaign Storage~\cite{campaign} and OceanStore~\cite{oceanstore} offer an
aggregate storage interface and are designed to provide storage and data access
facility to geo-distributed sites. 
However,
%these studies pose two limitations,
%i.e. no data sharing control is provided, data always need to be stored via
%aggregate storage interface and local data center file system cannot be used.
in these systems, users cannot selectively publish 
%the
datasets.
%\cb{(hyunki:can you mention example for data sharing options? I can't get what is data sharing options exactly.)}
More importantly,
shared datasets %data 
always need to be stored via the aggregate storage interface,
%and local %data center file systems cannot be used.
%\cred{Kim: Why is this (local file systems can't be used) problem? Remove this sentence.}
CFS~\cite{cooperative} allows the read-only access to shared data whereas, read-only access is not aligned with collaboration activities.
%\cb{(hyunki:which is not aligned with collaboration activities? read-only access or shared data?)}
%to use local data center file system. 
%Collaborators are required to use aggregate storage interface and also provide no data sharing control.
%span the globe 
%\cred{Kim: spanning the globe? it does not make sense to me. Can you rephrase it?}
%providing access to data stored from multiple sites, but lack\cred{\sout{s}} the ability to provide local write and read option
%\cred{Kim: Suddenly, "local write and read option" appears, but I do not know what you mean.}. 
 % requiring huge data copying and bulk transfers. 
The file system unification studies, including %includes 
WheelFS~\cite{wheelfs},
UnionFS~\cite{unionfs} and GBFS~\cite{gbfs}, %. All of these studies 
are focused on
providing a full-featured file system atop %of 
deployed file systems.
%however,
However,
they do not provide collaboration-oriented features, such as data sharing
control and advanced data discovery services.

Another important factor of the scientific collaboration is tight coupling
to POSIX interface.
%\cb{Taeuk : Actually, I could not find the connection between following two sentences. Most applications requires POSIX compatible API, but why DB requires massive data import?}
Traditionally, most scientific applications have been written to store and retrieve datasets
using POSIX-compatible file systems~\cite{vsfs}.
Introducing a new interface for the purpose, e.g., relational databases~\cite{scidb}, requires costly migration of existing datasets and 
unnecessary learning hassles to scientists.
%Most scientific applications require POSIX-compatible file
%system APIs to access datasets~\cite{vsfs}. 
%Using different storage options
%such as databases~\cite{scidb} might require massive data import. 
%Apart from it, 
In addition, scalable and efficient scientific discovery and search services,
e.g., extracting desired datasets from billions of file system entries,
are becoming an important component in HPC. 
%especially \cb{atop} %of \cb{the} file system namespace. 
%The desired dataset is filtered from millions and billions of files stored.
%Initially, such search feature is handled by having additional databases
%running atop of file systems. Later, the following studies such as
%VSFS~\cite{vsfs}, Klimatic~\cite{klimatic} and TagIt~\cite{tagit} integrated
%such service at the file system layer. 
Recent studies, such as 
VSFS~\cite{vsfs}, Klimatic~\cite{klimatic}, and TagIt~\cite{tagit}, integrated
such data management services at the file system layer,
instead of deploying additional database systems.
%Such discovery and search services 
Providing the data management services
are also important in collaboration environments,
because it can eliminate unnecessary data transfers between facilities
by quickly identifying and extracting datasets of interest.
\tbstore{} provides a virtual collaboration workspace to facilitate scientific collaborations.
The collaboration workspace provides common data visibility and also supports
the advanced data discovery services in a high-speed network connectivity.
%o
%. The POSIX-compliance frees the legacy applications from modifications. 
%It should offer 
It is crucial to present a single pathname to view and share a dataset, even when multiple data
centers or sites participate in the collaboration.  
%Additionally, 
Moreover,
%to facilitate effective and desired dataset retrieval in scientific collaborations, 
%to effectively retrieve desired datasets, 
%it 
the collaboration workspace should %offer 
support %discovery and search-query like features 
%the data management services, 
%the 
advanced data discovery services,
%\cb{Taeuk : data management service is mentioned just before. Isn't it about data discovery service?}
e.g., attribute-based file search queries,
to effectively retrieve desired datasets
and avoid unnecessary data transfers.
%\cred{Kim: suddenly data transfer overheads appear. It does not make sense to me.} 
%Atop, considering realistic scenarios, a scientist might be involved in multiple collaborations. 
In addition, it is common that a scientist participates in multiple collaborations~\cite{lim:2017:sc}.
%To accommodate such demand, none of the existing studies provide any additional feature such as support for multiple collaborations. 
To the best of our knowledge, none of existing systems directly support 
%the 
multiple collaborations, which we address via providing template namespace.
%\cb{(hyunki:so what? Does Scispace has a feature for that?)}
%However,
\tbstore{} offers a gluing POSIX-compliant interface atop %of 
dissimilar file systems %deployed 
%at 
from different geo-distributed HPC data centers. 

%% file: overview.tex
\section{\tbstore{}: Scientific Collaboration Workspace}
\label{sec:overview}
%In this section, we first present our key design goals and describe the architectural overview of proposed service framework for scientific collaborations.

In this section, we present our key design goals and discuss %in detail 
the design and implementation of \tbstore{} in detail.

\label{subsec:scifsoverview}
\begin{figure}[!t]
	\footnotesize
	\begin{center}
		\begin{tabular}{@{}c@{}c@{}}
			\includegraphics[width=0.48\textwidth]{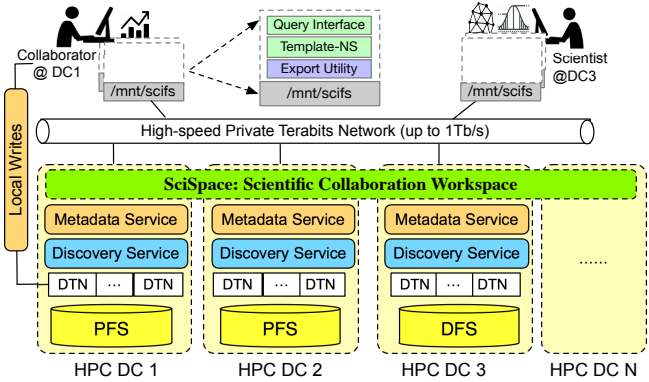}
		\end{tabular}
		\vspace{-0.1in}
		\caption{\small An Architectural Overview for \tbstore{}. }
		\label{fig:archi_overview}
		\vspace{-0.3in}
	\end{center}
	%\vspace{-0.in}
\end{figure}

%describe the design and implementation of each collaboration workspace framework component.

 \begin{comment}
\cb{Lets discuss about the overview and add new overview figure unlike istore. I think just write few lines about every module  may be using itemized and referring to the figure. i) Don't forget to mention about providing data access to any site with single uniform path name ii) single place to locate information stored by multiple sites in design goal. it will be my design goal. easy mount point integration on top of existing posix-complaint single and distributed file systems. easily portable, low complexity as compared to other fs architectures}.
\end{comment}

\vspace{-0.1in}
\subsection{Goals}

%\cb{\sout{We list our key design principles in this subsection.}}

\squishlist
\item
{\bf{Collaboration Workspace:}}
%\bf{Collaboration-Friendly Workspace:}} 
The key design goal is to provide consolidated data visibility to all collaboration data centers
%% hs: i am confused, what the 'path' specifically means?
under a single uniform namespace. %path.
A %single 
workspace is layered atop multiple dissimilar %dis-similar 
file systems mounted on %via 
data transfer nodes,
%showing 
and presents
a common unified data view to all participants in %scientific collaborations. 
the collaboration.

\item {\bf{Native-Data Access Support:}} 
To keep minimal modifications while achieving high performance, we consider it important 
to %offer the 
support for local writes and %local 
reads %to collaborators 
using local data center's file system namespace. 
%% hs: the following sentence needs clarification.
%\cred{We consider keeping autonomity across the collaboration sites. }
%\cred{Kim: what about local-reads? Are not "local read and write" all "native data access"?}

\item
{\bfseries{Multi-Namespace and Selective Data-Sharing:}} 
In real-world scenarios, it is common that a single scientist %collaborator 
is involved in multiple collaborations. 
Moreover, offering the ability to
selectively share data via different namespaces for each collaborator.
%% hs: I guess we need to mention namespace at least once in this paragraph?
%% the title is "multi-namespace"
Thus, we added privilege in our design to manage multiple collaboration workspaces.

\item
{%\bfseries{Efficient Scientific Discovery \&  Search:}} 
\bfseries{Efficient Data Discovery and Search:}}
In geo-distributed collaborations, the extraction of required and useful data
is of high significance. Additional performance overhead and network cost can
be incurred if the required dataset is not intelligently retrieved. To
incorporate such intelligence, we consider the scientific discovery and search
service as an important design goal. 
%We intend to provide a query-like multi-criteria search facility.
\tbstore{} supports attribute-based data search facility.

%\item
%{\bfseries{Efficient Metadata Management:}} The simple file metadata and complex metadata such as HDF5 attributes are extensively studied in the scientific domains to speed up the scientific applications. Our design includes distributed metadata management where, each data transfer node holds a DB-shard (a partition or instance of a database running at each data transfer node), to avoid the I/O contention incurred by centrally managed metadata.

\begin{comment}
\item
{\bfseries{Policy-driven Load-Balancing:}} The load-balancing plays a vital role in distributed and multi-node environments, as maximum throughput can be achieved, if the system is fairly load-balanced depending on priorities and requirement, i.e., storage and computation. The placement policies are important in geo-distributed environments, hence we considered it as a tunable parameter which can be configured at system deployment time.
\cred{Kim: I am not sure if you implemented this goal. If not, drop it. If my understanding is wrong, pls correct me.}
\end{comment}

\squishend

%% file: design.tex
\vspace{-0.05in}
\subsection{Scientific  Collaboration  Workspace}
\label{sec:design}
\vspace{-0.05in}
%\cb{\sout{The architectural overview of the proposed collaboration framework is presented in Figure~\ref{fig:archi_overview}.}}

The proposed collaboration model renders a global picture of shared data to all the participants in the collaboration. An architectural overview of the proposed collaboration workspace is shown in Figure~\ref{fig:archi_overview}.
%Our collaboration workspace follows the concept; \textit{what contents irrespective of where located}. 

%Next, we describe the design of components comprising the proposed collaboration workspace.
 
%Collaborators can view and access data shared by multiple data centers under a single path. The key point of interest here is; \textit{what contents irrespective of where located}.

%In this section, we discuss in detail the design and implementation of key \tbstore{} modules. 

%First, we describe the global namespace, then distributed metadata and data discovery service is implemented and integrated in \tbstore{} and lastly, we shed light on metadata export utilitiy and load balancing.
%\vspace{-0.1in}

\subsubsection{Unified Virtual File System Layer}
\label{sec:uvfs}
%% hs: I think we need to careful when we say POSIX here.
%POSIX semantics and POSIX interface are two different things.
%The below says that providing POSIX semantics is important. After that it says
%\tbstore{} provides POSIX-like file system API.
%I do not understand: 
%Does the \tbstore{} file system fully follow POSIX semantics? (e.g., NFS is not
%POSIX but relaxes the semantics by close-to-open.)
%Merely supporting POSIX open, close, read, write, ... I/O syscalls does not mean
%it is POSIX-compliant.

The Scientific Collaboration Workspace empowers \tbstore{} to elude the need for modifications to existing scientific applications and file system architecture. The intention to keep the existing application and storage architecture intact drives the need to implement a file system 
%like 
interface which can offer POSIX semantics. Besides, all collaboration participating geo-dispersed data centers grants access to shared resources such as storage and compute nodes via single or multiple DTNs. The effective utilization of provided multiple DTNs is also an essential viewpoint which needs to be considered. If not properly approached, it can lead to bottlenecks, i.e.,  multiple collaborators accessing a single DTN. To this end, our Scientific Collaboration Workspace is equipped with a POSIX-like file system API and provides all the basic file system operations. To manage the metadata effectively, we employ a distributed metadata architecture and details are presented in next subsection~\ref{sec:metadatadb}. 

\begin{figure}[!t]
	\begin{center}
		\begin{tabular}{@{}c@{}c@{}}
			\includegraphics[width=0.48\textwidth]{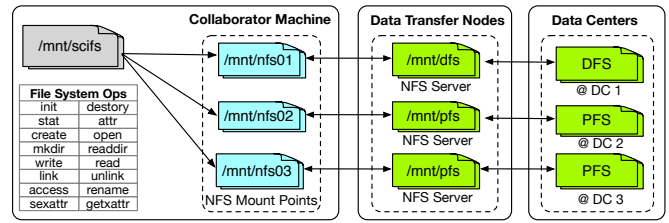}
		\end{tabular}
		\vspace{-0.1in}
		\caption{\small Scientific Collaboration Workspace.}
		\label{fig:namespace}
		\vspace{-0.25in}
	\end{center}
\end{figure}

An important role of Scientific Collaboration Workspace includes providing a
consolidated view of shared data dispersed across dissimilar file systems
deployed at geo-distributed data centers. 
Figure~\ref{fig:namespace} shows how \textit{scifs} is mounted on the collaborator's machine.
The participating data centers accordingly grant the access from the collaborator's machine
through DTNs.
Compared to the traditional approach,
where scientists have to manually transfer data between multiple DTN mountpoints,
\textit{scifs} mountpoint (\textit{/mnt/scifs}) provides a seamless integration of multiple mountpoints
and user-transparent data transfers.
%describes a case where, \textit{scifs}, file system interface of \tbstore{} is mounted
%onto the collaborator machine. 
%All geo-distributed data centers provide
%privilege to deployed file systems via DTNs shown in
%Figure~\ref{fig:namespace}. 
%A naive baseline solution includes mounting all DTNs to collaborator machines. However, this approach neither provides global data view nor promote effective data sharing. To share a single file, collaborator needs to copy the file to all the NFS mount points manually, whereas, in real scenarios it is difficult and time-consuming. 
%\cred{Kim: Can you rephrase this sentence? I don't understand what you are saying from "witch in real...}
%In \tbstore{}, 
%Scientific Collaboration Workspace is capable to glue the NFS mount points under a single path as shown in Figure~\ref{fig:namespace}. It provides a virtual abstraction on top of NFS mount points. %At mount time, all NFS mount points are input to \tbstore{} to create an umbrella on top of all NFS mount points. 
More importantly, the \textit{scifs} mountpoint acts as a visible and
interactive collaboration workspace, as traditional file systems do, where all
standard file operations take place.
%All file system operations are entertained via this collaboration workspace. 
When an incoming write request is received, Scientific Collaboration Workspace assigns a DTN for the write request by hashing the file pathname.
%hashes the file path and routes the request to an assigned DTN via
%hash-computation. 
\tbstore{} internally maintains a distributed metadata database
to store all file metadata including the computed hash value (\autoref{sec:metadatadb}). We used hash-based placement strategy in order to eliminate I/O broadcast problem when multiple DTNs host metadata service.

%The computed hash value, file name and path metadata is
%stored in internally in \tbstore{} for future reference.  In particular,
%\tbstore{} maintains databases internally, the design and implementation of
%which will be explained in next subsection.

%File Mapping schema as shown in Figure~\ref{fig:shard}. 
%\cred{Kim: Suddenly file mapping schema is referenced without explaining what it is before. Perhaps, you can say it is stored internally in \tbstore{} for future reference.}
%\cred{%%hs: I do not understand the following sentence.
%The reason to use the hashing for placement is to reduce unnecessary metadata contact points between workspace and collaborators, in particular, for metadata-intensive operations.  }
%along with efficiently load-balancing the DTNs.

When a collaborator wants to read a specific file, the hash is computed against the file pathname and a request is sent to an appropriate DTN hosting the file metadata information. The directory listing file system functionality (such as ls) provides a list of shared contents to the collaborator by fetching file metadata information from all the DTNs in a parallel fashion.
%list the shared contents/files inside \tbstore{}, the directory listing operation (ls) entertains such request. At first, all NFS mount points (data center file systems) information is retrieved 
%from metadata service  \cred{Kim: again metadata service is not defined what it means in \tbstore{}.}
%and then, it
%\cred{Kim: we? or it?} 
%broadcasts parallel requests to all the DTNs for providing the list of files. %in File Mapping schema. \cred{Kim: file mapping schema is not defined yet.}
%In addition, collaborators may want to publish their datasets selectively to the global collaboration workspace.
%At this point, collaborators can show concerns about sharing file or data directly to global collaboration workspace. 
%To address such sharing concerns, 
Such design provides ease in controlling data sharing semantics. The collaborators can selectively publish datasets in the collaboration workspace.  
We maintain a flag \textit{sync} as an extended attribute of each file. When files are stored directly via \tbstore{}\textquotesingle s workspace, the flag with value sync = \textit{true} is added. The {ls} operation lists only the files and directories with the sync flag set true. In the current implementation, \tbstore{} do not offer file or data removal to remote collaborators, but it can be easily extended via the metadata service.

\subsubsection{Metadata Management} 
\label{sec:metadatadb}
%Metadata is of high significance in any file system because it is the key input to all operations.

%Metadata is of high significance in the file system because it is the key input to all services. The need for efficient metadata increases as the number of participating collaborators increases. Moreover, to effectively speed-up scientific workflow, we choose distributed metadata architecture. 
Metadata is of high significance in file systems because it is the key input to all file system services. In collaboration environments, the need to minimize the metadata bottleneck originates when collaborator traffic increases. %The collaborators might run metadata-intensive applications accessing data shared across geo-sapatial sites connected via high-speed network. 
We adopted distributed metadata to reduce metadata bottlenecks caused by the central metadata management approach. 
%%
%% hs: I do not think the following is true.
%% maybe we can tone down like:
%% \cb{In addition, the distributed metadata management architecture also helps to provide efficient
%% scientific indexing and search services.}
%%
Distributed metadata provides more efficient scientific search
and indexing services than a centralized indexing approach.

% We used DTNs to implement distributed metadata services and we explain why we choose DTNs to run the dist. metadata services
%We implemented a distributed metadata system that uses DTNs from each data center for implementation. 
The metadata service in \tbstore{} is running on every DTN from all participating data centers.
The reason to execute metadata services on DTNs is manifolds, (i) we can effectively utilize the DTNs, (ii) storing metadata globally enables us to provide metadata to all the collaborators mounting \tbstore{}, and (iii) we can exploit multiple available DTNs as distributed metadata services for efficient scientific discovery and indexing as compared to centralized metadata approach.  %The file system metadata such as filename, size, owner and path are synchronously update when a request is received.
% Now we explain about DB sharding and its benefit. 
To keep our design scalable, we split %the database maintaining 
metadata into multiple partitions. This partitioning helps in obtaining a fair load-distribution across available DTNs. 
Each %running 
instance of metadata partition acts as a DB-Shard (database shard).

\begin{figure}[!t]
	\begin{center}
		\begin{tabular}{@{}c@{}c@{}}
			\includegraphics[width=0.48\textwidth]{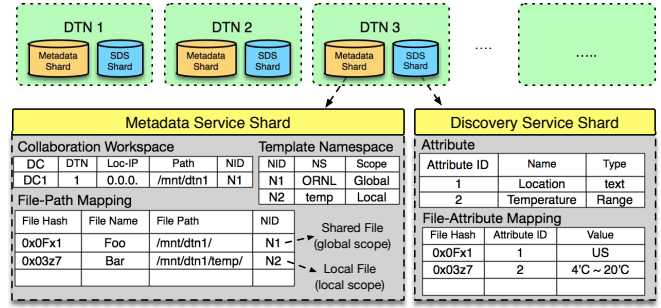}
		\end{tabular}
		\vspace{-0.1in}
		\caption{\small A schema view for Metadata and Discovery Shard.
	%	\cred{Kim: this figure drawing should be well aligned with Fig2 drawing.}
		}
		\label{fig:shard}
		\vspace{-0.3in}
	\end{center}
\end{figure}

%With this DB-shard architecture, 
Specifically,
each DTN maintains %runs 
two DB shards, i.e.,  metadata service shard and discovery service shard,
as shown in Figure~\ref{fig:shard}.
%two shards as shown in Figure~\ref{fig:shard}, i.e., {\em metadata service shard} and {\em discovery service shard}. 
We maintain two different types of metadata, i.e., file system metadata and indexing metadata. 
The file system metadata, such as filename, size, owner, and the path,
is synchronously updated when a write request is received. The indexing metadata includes metadata of scientific dataset headers (such as HDF5 and NetCDF self-contained attributes) and user-defined indexing attributes.
%However, for index metadata, 
For index metadata,
%\cred{Kim: you explained what the file system metadata is but, you didn't what index metadata is. Briefly explain what they are and mention details of how they are used will be explained in next Sec 5(?)}
we provide both synchronous and asynchronous DB update mechanisms. 
In synchronous DB update, the file indexing and metadata extraction %is done
is performed
when a write request is received. 
%%hs: do we show this in the evaluation?? otherwise we need to reword it.
It incurs high overhead but it can be masked under FUSE layer overhead.
Whereas, in asynchronous DB update, the file indexing and metadata extraction is conducted later after file is stored. 
Only a single message is sent to indexing service to register the file for indexing and metadata extraction. 
When to conduct the indexing and metadata extraction depends on pre-defined threshold such as time, size and file count.  %a single message is sent to indexing service when write request is received to the add the file name for indexing in future.  
%\cred{Kim: this setence is broken.}
%The asynchronous DB update is triggered based on multiple thresholds 
The asynchronous DB update exhibits inconsistency between the file system metadata and 
the indexing metadata, depending on how early the metadata extraction and indexing is performed after
the corresponding file operation.
We further explain the pros and cons of two DB update mechanisms in Section~\ref{sec:discovery}.
This distributed metadata architecture is tightly coupled within the
collaboration workspace. We adopt an index data structure to promote effective
lookup and search queries on top of relational database to enable file
attribute based retrieval. 
We do not use key-value stores, as our metadata indexing approach requires
multiple associations, e.g., %. For example, 
linking a single file with multiple
attributes or single atttribute to multiple files. % to a single file. %, 
%which we believe limits the lookup and query operations in collaborations. 
The schema for collaboration and indexing metadata is shown in Figure~\ref{fig:shard}.
%% this sentence has been moved.
%Whereas, 
Note that
such attribute-based file retrieval is not
possible in the traditional approach
without performing a costly exhaustive search.
\tbstore{} obtains two significant benefits by integrating file indexing and attribute extraction at file system layer;
%\cred{Kim: metadata indexing... this is not clear to me. Can you name it better? Something like to capture the meaning that the distributed metadata system is tightly coupled within the file system.}
(i) effective execution of metadata-intensive I/O operations such as file name
and path mappings on specific data center,
(ii) no crawling/file lookup required on %top of multiple dis-similar file system namespaces,
multiple file system namespaces,
(iii) empowering search and query based on custom-defined attributes, file system
stat attributes and scientific dataset attributes (such as HDF5 self-contained
attributes). 

\subsubsection{Local-Writes and Export Protocol}
\label{sec:metadata}

%As the Scientific Collaboration Workspace provides the file system functionality, 
%\tbstore{} can write data to both local and remote data centers via collaboration workspace.
The file system interface of \tbstore{} (\textit{scifs}) allows collaborators to seamlessly access 
local and remote datasets in the collaboration workspace.
However, %it incurs a certain performance degradation due to our prototype implementation via FUSE API. 
the additional file system layer, written using the FUSE framework in our prototype (Section~\ref{sec:uvfs})
may degrade the overall I/O performance.
%We believe that such performance degradation can be overcome by providing
%support of local-writes, 
To avoid such performance degradation, \tbstore{} supports local-writes,
i.e., writing data directly to the local data center
file system %namespace bypassing 
instead of the
collaboration workspace (FUSE layer).
%For example, if a collaborator is privileged to store data directly to local data center namespace engaged in collaboration, then high performance can be obtained.
Through the local-writes, \tbstore{} can deliver the native performance of the local
data center file systems when collaborators can exploit the local file systems.
%\cred{Kim: Isn't this sentence redundant with the preceding sentence?}
%\cred{Kim: Suddenly, I was wondering if there was an explanation before, when writing to a scientific workspace, it will write to a local PFS (not a remote PFS). It seems to be somewhat confusing to this case with the local-write mentioned here.}
%However, not supporting local namespace usage leads to severe performance degradation due to FUSE layer and additional metadata management overhead. 
%\cred{Kim: This sentence has already been explained similarly.} 
%The benefits to allow local-writes includes no additional consistency and
%fault-tolerance management, low network traffic across the collaborating sites
%and zero-performance overhead. 
Furthermore, the local-writes also reduce the network traffic across the sites and simplify the consistency and resilience managements due to direct storage at local data center namespace. %% hs: why??? it looks like it will be more complicated?
%The current Collaboration Workspace can list
%only the contents/files available at each DTN metadata shards. In order to
%export contents to the collaboration workspace, it requires the metadata of the
%contents/files stored via local file system namespace. However, exporting such
%data to collaboration workspace is challenging. The challenge might include how
%to synchronize such data to the collaboration workspace.
However, datasets that have been written through the local-writes
are not directly visible inside the collaboration workspace, and thus
should be properly propagated to the file system namespace of the collaboration workspace.

%To assist local-writes support in \tbstore{}, we propose to build \textit{Metadata Export Utility} (MEU). 
To assist local-writes, \tbstore{} features \textit{Metadata Export Utility} (MEU), which commits all unsynchronized metadata of locally-written datasets to the file system namespace of the collaboration workspace.
%\cred{\sout{The motivation behind is to address the data sharing needs in collaborations.}}
%It enables \tbstore{} to allow the collaborators to use local data center namespace. 
%Later, commit all the unsynced metadata of contents/files to the collaboration workspace.
%This commit can be carried on-demand. %or can be scheduled %via cronjob scheduler
In addition, collaborators can explicitly trigger such commits.
%\cred{Kim: what is the cronjob scheduler?}. 
This concept works in a similar fashion to git local and remote repository management. 
%\cred{\sout{
%% hs: i don't think we need to introduce git this kindly.
%Git is an open-source distributed version control system and offers local
%staging area for keeping local changes.
%As user commit the changes, all the changes are pushed to the remote repository.  However, in}}
%% hs; namespace doesn't store the data
In our design, %the local namespace stores the data 
because datasets written via the local-write are stored
in permanent storage %repository 
(local data-center file system) %namespace) and 
%requires the metadata synchronization with \tbstore{}.
only their metadata needs to be synchronized with the collaboration workspace namespace.
%We design the export utility for a versatile functionality.  It 
MEU appropriately synchronizes such metadata into the collaboration workspace namespace. %not only allows the local writes support 
%but also 
In addition, MEU allows a fine-grained control for sharing the datasets,
e.g., when a collaborator wants to share only subset of a dataset via collaboration workspace.
%provides the control on data sharing attributes. 
%For example, if a collaborator wants to disable sharing on a particular file or
%group of files stored via collaboration workspace or wants to share/export the
%data stored via local namespace, the export utility can assist in such
%situations. 
%\cred{Kim: Suddenly motivation comes here. Should it be moved to come earlier above?}

\begin{figure}[!t]
	\begin{center}
		\begin{tabular}{@{}c@{}c@{}}
			\includegraphics[width=0.50\textwidth]{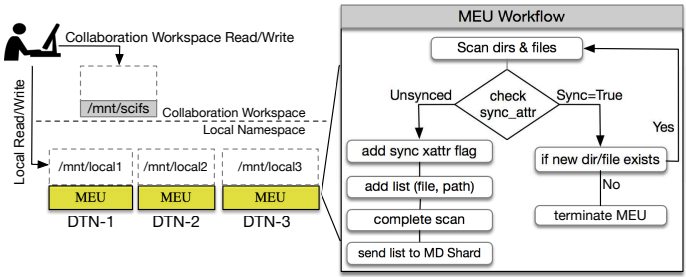}
		\end{tabular}
		\vspace{-0.1in}
		\caption{\small Local-Writes and Export Protocol.}
		\label{fig:meu}
		\vspace{-0.3in}
	\end{center}
\end{figure}

The local-write and MEU workflow is 
shown in Figure~\ref{fig:meu}. %It 
MEU scans the files and directories recursively
from a certain local directory, such as /home/project.
%the root point (such as /mnt or /home in Linux) at each local data center namespace. 
During the scan, %phase, 
it checks the extended attribute \textit{sync} of each file and directory in a pathname. %at each directory and file level hierarchy.
For example, to examine /foo/bar/hello.hdf5, %the 
MEU %will 
first checks the extended attribute of \textit{foo}.
If the flag is true, %then we skip this directory depth because of already synced metadata, else 
MEU skips the entire directory because all files and directories under \textit{foo} have already been synchronized.
Otherwise, MEU enters the directory and scans entries.
%we go deeper in the directory and repeat the same.
Whenever any change occurs inside a directory, we modify the flag of %atop of 
the %immediate 
parent directory of the file or directory (in the example, \textit{bar} is the %an
%immediate 
parent of \textit{hello.hdf5}). Once the scan phase finishes, we add an
extended attribute to all unsynchronized %un-synced 
files. %, and synchronize them with \tbstore{}
%metadata entries via a single message packing all the un-synced files metadata.
When MEU synchronizes the metadata, it packs all unsynchronized metadata into a single message
to minimize the synchronization overhead.

\subsubsection{Template Namespace}

%\vspace{-0.03in}\

%\cred{ Is not template namespace capable because SciSpace has metadata in DB? The contents of Fig4 should be explained first. Because DB-shard manages distributed metadata and performs indexing, it is natural that the template namespace can be implemented.}

\tbstore{} is intended to effectively satisfy the needs %of collaborations. 
for various types of collaborations.
%\cred{Kim: delete this sentence.}
%\cred{\sout{Though in practical scenarios, two observations cannot be disregarded, i.e.,  i) a collaborator may require individual workspace for own research, simulation, and analytical jobs, and ii) a single collaborator may be involved in multiple collaborations simultaneously.}} 
For instance, a %A 
collaborator may require a dedicated %individual 
workspace for own research, simulation,
and analytical jobs. %, or 
Also, a %single 
collaborator may be involved in multiple collaborations simultaneously.
%For example, cloud 
Cloud data storage systems such as Dropbox and Google Drive 
permits sharing data with multiple users, and %multiple users 
a user can participated %overlap 
in multiple projects and collaborations.
%\cred{Kim: This example is unnecessary.}
Based on 
%the observations, 
these practical use-cases, %scenarios,
%we approach such needs by sketching 
\tbstore{} provides
a namespace management module, \textit{Template Namespace},
based on the distributed metadata management architecture.
%% hs: what does this mean? please improve to include this sentence.
%\cred{with the same notion of the template.} 
%\cred{The distributed metadata privilege us to provide such a feature required in collaborations.}
Collaborators can define multiple namespaces in \tbstore{} 
%% hs: I dont understand the following expression.
with the scope of each namespace (local/global).
Figure~\ref{fig:shard} %clearly 
shows the association between Template Namespace and other metadata in \tbstore{}.
%of Template Namespace with all \tbstore{}\textquotesingle s metadata relations. 
In specific, when a file is written, its pathname
%When some content/file is stored, the path 
determines the namespace, which in turns defines the scope of %that content. 
the file content.
If a namespace scope of a file is local, the file is only visible to the owner of the file.
%then contents/files are not visible/shared with any of the collaborators except the owner of the file, else
Similarly,
if the scope is global, %then, contents/files are visible at remote sites.
the file becomes visible to any collaborators within the collaboration workspace, e.g., a remote collaborator.
%\cred{\sout{Currently, template namespace is in the development phase.}}

%Each namespace carries its own properties and each feature can be enabled or disabled by the collaborator, e.g., if collaborator does not want namespace to export data to the collaboration workspace, he can switch off the export attribute. All such policies are managed via metadata manager. The metadata manager has namespace schema, where all the attributes of each namespace are managed. When any data discovery service is invoked, first metadata manager is requested to provide namespace configurations, then the further action is taken based on configured policies. 

%similar to C++ programming language template concept, as \textit{Template Namespace}. 
%Each namespace carries its own properties and each feature can be enabled or disabled by the collaborator, e.g., if collaborator does not want namespace to export data to the collaboration workspace, he can switch off the export attribute. All such policies are managed via metadata manager. The metadata manager has namespace schema, where all the attributes of each namespace are managed. When any data discovery service is invoked, first metadata manager is requested to provide namespace configurations, then the further action is taken based on configured policies. 

%\cred{The template namespace is possible because the idea illustrated in Figure 4 is implemented.}

%Currently, template namespace module is in the development phase. We are improving and aligning our template concept near to the real scenarios.

\subsubsection{Scientific Discovery Service}
\label{sec:discovery}

%\cred{\sout{In this section, we describe the rationale behind the design and implementation of Scientific Discovery Service (SDS).}}

%%hs: do we really need the following sentence??
%\tbstore{} offers Scientific Discovery Service (SDS).
%To extract the desired data among millions and 
Extracting a desired dataset from
billions of data files remains a
central interest of the science and research communities. 
%The motivations to adhere Scientific Discovery Service (SDS) in Collaboration
%Workspace includes the following;  
Particularly, a support of Scientific Discovery Service (SDS) 
within the collaboration workspace provides the following benefits;
i) it frees %to free 
collaborators from retrieving undesired data to
local data centers via data transfer tools such as bbcp~\cite{bbcp},
LADS~\cite{Kim:2015:LOD:2750482.2750488}, and ii) %\cred{\sout{circumventing}}
it circumvents %to circumvent 
%% hs: the following sentence needs to be improved.
%\cred{
	manual dataset screening  %on local data center for  %phase screening
	 phase in scientific workflows performed before analysis. %}
%%The recent TagIT~\cite{tagit} study provides indexing and discovery service in GlusterFS,
%however tight coupling with Glustre architecture limits the use of TagIT in our
%collaboration model. However, in the stated design, we consider it as a
%separate flexible module and keep the indexing metadata apart from
%collaboration workspace metadata as shown in Figure~\ref{fig:shard}. This
%design choice empowers \tbstore{} with modular flexibility such that, indexing
%service can be skipped when not required. 
%
%% hs: I don't think we need to break the paragraph
%As 
However, since the
SDS service entails additional processing for generating per file indexes,
it may incur %incurs certain overhead. 
a certain performance overhead.
%Whereas, 
%\cred{Kim: this sentence is broken}
%in some cases, 
Therefore, if such an indexing is not required on a certain dataset,
it is %might be 
favorable to skip the indexing for the dataset
%, when not required, 
to avoid the overhead. 
For instance, %example, 
an application %access pattern requires only storage and no immediate analysis.
may only require a storage space without having any subsequent analysis tasks.
In addition, it is possible that a scientist does not need such an indexing feature
for a certain dataset.
%Another pattern may follow both storage and
%immediate analysis or collaborator is not interested to generate indexes
%against specific files. 
%To cater such cases, we consider worthy to design
%versatile metadata extraction modes to index data in SDS service. 
To support such various requirements, \tbstore{} provides three different metadata extraction modes.
%% hs: all the following sentences look unnecessary. 
%% what are the "several benefits"?? if not discussed here, we do not need it here.
%Additionally, offering multiple modes can give \tbstore{} several benefits in collaboration.
%% hs: this too narrows the benefits of having the metadata indexing. it should be discussed
%% in the beginning of the section, when we explain the motivation.
%These metadata extraction modes are useful when analysis is performed after
%data generation.
%\cred{Kim: Did not you explain that you use SDS when you explain the MEU?}
%\cb{Khan: I think i did not explain SDS in MEU section.}
%\cred{ If so and you want to keep that story, SDS should have been mentioned there that its details will be explained later.}
%\cb{Professor; sorry i could not follow your point?}

%Next, we describe all the proposed metadata extraction modes.
%\cred{Kim: here a term of indexing seems to not be proper. }
%\cred{Kim: basically three modes described below are how metadata extractions are performed, whether in sync, async, or after all data generations are all done, offline. In the above paragraph, why don't you say the versatile modes are basically useful when analysis is performed after data generation. For this, when to extract metadata (files attributes) are critical because it affects analysis performance.}

%% hs: redundant
%Thus, we support three different modes for metadata extraction.

\begin{figure}[!t]
	\begin{center}
		\begin{tabular}{@{}c@{}c@{}}
			\includegraphics[width=0.48\textwidth]{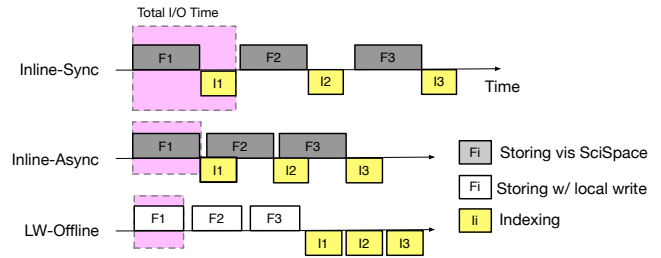}
		\end{tabular}
		\vspace{-0.1in}
		\caption{\small \tbstore{}: Metadata Extraction Modes.} 
		\label{fig:indexing}
	\end{center}
	\vspace{-0.3in}
\end{figure}

\squishlist
%\textbf{
\item
Inline-Sync: In this mode, write operation includes both data storage and
metadata extraction in a synchronous way. 
As depicted in Figure~\ref{fig:indexing}, a write operation completes
only after all the metadata is extracted and indexed.
%Until all the data is indexed, write
%operation is not finished as shown in Figure~\ref{fig:indexing}. 
%We propose this %approach 
%for collaborations with access pattern requiring both storage and
This mode aims to facilitate applications that require both storage space and
immediate analysis on produced datasets. 
%This mode increases the I/O latency and degrades I/Os per second due to inline metadata extraction overhead.
Although the Inline-Sync mode provides a strict consistency between datasets and the index database,
its synchronous metadata can significantly slowdown the individual I/O operations.
%\cred{Kim: Pls improve Fig6. Left and right spaces of the drawings are wasted. You may draw it wide.}
\item
Inline-ASync: To reduce the increased I/O wait time, we propose
Inline-Async mode, that injects partial de-coupling between storage and
extraction operation. As shown in Figure~\ref{fig:indexing}, the total file write
time in Inline-Async does not include the metadata extraction process. %, unlike Inline-Sync approach. 
%We accomplish so by adopting 
In specific, we adopt
a queue-based metadata extraction
architecture, where %whereby 
an indexing request message is enqueued when a file is written. %to SDS and 
SDS asynchronously dequeue messages and index data accordingly. This 
%approach can be of great help when deployed in 
mode specifically targets
environments with offline or delayed analysis after data generation. %storage.
%% hs: the previous Inline-Sync doesn't say anything
%% about the implementation.
It includes FUSE and negligible message enqueue overhead.
\item
%\textbf{
LW-Offline: 
%\cred{Kim: What does LW stand for? Why don't you simply call offline versus inline?}
To support indexing on top of local-writes (LW), we require offline indexing mode
which directly performs the metadata extraction within the %executes on 
data center file system namespace.
%It is required in cases 
This mode aims to faciliate cases
when datasets are %is 
stored via the local namespace and %or 
high-performance %gain 
is expected.
The indexing service is triggered on the DTN directly. The write
operation includes no FUSE overhead due to native access. 
\squishend

%{\em Use-cases:}
In the scientific community, the HDF5 and NetCDF datasets are most commonly used data formats~\cite{hdf5}. %HDF5 file is specially designed for unstructured data where a single file can act as big file comprising multiple small files~\cite{hdf5}. 
%To index HDF5 files, 
SDS utilizes the HDF5 library~\cite{hdf5} to extract all the attributes from
the HDF5 file. 
%\cred{%% hs: I do not understand the following. how is it different from the user-defined tagging?
Collaborators can specify attributes to index data in \tbstore{}.
The SDS validates the data for matching attributes defined by the
collaborator.
If the match is found, the entry (attribute, file, value) is
recorded in the Discovery shard as shown in Figure~\ref{fig:shard}. In
addition, we also offer manual or collaborator-defined %desired 
tagging, where a
collaborator is facilitated to tag file or group of files with custom attributes. %a custom
%% hs: what's the special cases?
%attribute. %, which might be required in some special case. 
The simple attribute
structure consists of \textit{attribute.name} which refers to attribute name,
\textit{attribute.type} refers to attribute datatypes, and
\textit{attribute.value} refers to the value of an attribute. In the scope of
current work, we provide only three types of attribute types,
i.e., integer numbers, floating point numbers, and texts.
%i.e., int, float, and text.
%However, we 
We plan to %are considering 
to extend our implementation to include range-based attribute datatypes. %% hs: what is the range-based attribute data types??

\begin{comment}
The indexing service reads the HDF5 file metadata which is different from the file system metadata. We check the user-defined parameters and match one-to-one all the attributes with HDF5 provided metadata. We push all the matching attribute entries to the database running at the local data center in order to keep tagging metadata near to the actual file location. In manual-tagging, we let all the files to be stored without appending additional overhead and then, manually trigger the tagging service, which reads the HDF5 file metadata and records all the matching attributes in the same fashion with auto-tagging. 
The motivation behind to provide such two different approaches to tag files is the HPC application diversity. In simple, applications can be categorized into two types. n such cases, the analysis is performed in later stages or may be conducted against legacy data. \tbstore{} offers the auto-tagging for the first type of applications, which fall into end-to-end category and manual-tagging can benefit the applications more focused on running analysis in later stages. All the communication between data discovery service and metadata manager is defined in standard messaging format using Google Protocol Buffer~\cite{ProtoBuf}. 
\end{comment}

%\textbf{
%{\em Search and Querying Interface (SQI):} 
%We provide 
%% hs: does the command line utility has a name? its better to mention the name.
%% instead of SQI
\tbstore{} provides a query interface via a command line utility. %in \tbstore{}. 
Using query interface, %% hs: SQI has never been discussed/explained previously.
%the 
collaborators can easily query %the Collaboration Workspace for 
the desired contents/files within the collaboration workspace.
%We interpret the different 
The command line utility supports
operators inside a query string, %when used in search or query string 
such as equal (=), greater (\textgreater), and less (\textless). %(=, \textgreater, \textless). 
For the text datatype, we provide equal (=) and like (\textit{like}) operation.

\tbstore{} currently delegates
%Currently, 
the fault-tolerance, replication, and data consistency %other data consistencies 
managements to distributed and parallel file systems inside data centers.
%are handled by data center deployed distributed and parallel file systems.
In fact,
\tbstore{} inherits all these features from data center equipped file systems,
because %we provide 
it merely adds a thin virtual abstraction layer %only virtual abstraction 
on top of the mountpoints of such file systems. %NFS mount points. 
%\cred{\sout{
However, we consider the collaboration workspace metadata replication 
as an important factor and plan to support the metadata %extend our prototype to include 
replication in future.

%% file: eval.tex
\section{Evaluation}
\label{sec:eval}

\begin{table}[!t]
\begin{center}
		\footnotesize
%	\caption*{\textbf{Lustre File System Configurations}}
	\caption{\small Description of Evaluation Test-bed Setup.}
		\vspace{-0.1in}
	\begin{tabular}{|l|l|}
		\hline
		\textbf{Component} & \textbf{Description} \\ \hline \hline
		Collaboration & 2 Data Centers \\ \hline
		Storage & Lustre PFS for each Data center\\\hline
		Lustre  & 4 Nodes  (2 x MDS, 2 x OSS) \\ \hline
		\multirow{2}*{MDS}  & 24 Intel Xeon E5-2650 CPU Cores, \\ & RAM 128 GB,  1 x 6.3 TB MDT 
		 \\ \hline
		\multirow{2}*{OSS} & 16  Intel Xeon E5-2650 CPU Cores,  \\ & RAM 64 GB, 11 x 7.2 TB RAID-0 OSTs
		\\ \hline
		DTNs & 4 Nodes (Lustre Client Nodes)
		\\ \hline
		Collaborators & 1-24 Collaborators 
		\\ \hline
		CPU Cores  & 24 x Intel Xeon E5-2650 @2.20GHz  \\
		Memory      & 128 GB \\
	%	Disk Storage      &   256 GB  \\	
		Network &  Infiniband EDR (100Gbps) \\
		OS     & CentOS Release 7.3 Kernel v3.10 \\	 
	%	File Systems & Ext4, Linux NFS v4, UnionFS, Lustre v2.9.0 \\ 
	\hline
%		Benchmark & IOR \\ \hline
%		\multirow{2}*{Real Dataset}  & MODIS-Aqua Level2 Ocean Surface Data\\ %& 116 GB HDF5 Dataset,  4600 files \\ \hline 
%		HDF5 Tools & H5Diff, H5Dump \\  
	%	\hline
	\end{tabular}
	\vspace{-0.3in}
	\label{tab:testbed}
	\end{center}
\end{table}

%We evaluate the performance of the \tbstore{} prototype using scientific datasets and representative workloads. 

%In the experiments, we deploy \tbstore{} on top of two geo-distributed data centers equipped with Lustre~\cite{lustre} to assess how efficiently the proposed collaboration workspace performs in the real collaborations.

%collaboration file system across two geo-distributed data centers. 
%First, we perform synthetic benchmarks for the \tbstore. Second, we choose from a collection of benchmarks that measure the behavior of the storage system under various workloads. We also evaluate the \tbstore{} metadata manager using benchmark tool to analyze the metadata overhead.

\subsection{Implementation}
\vspace{-0.05in}
%\subsubsection{Collaboration Workspace}
We implemented \tbstore{} using the FUSE\textquotesingle s high-level API v2.9.4~\cite{tofuse}. %\cred{\sout{File System in Userspace (FUSE) is the most widely used framework to prototype and evaluate new approaches to file system design~\cite{tofuse}.}}
Our implementation fully complies with POSIX standards and shows UNIX-like semantics and directory structure. A generic messaging protocol is employed to interact with all the components of \tbstore{}, 
accomplished via 
Google Protocol Buffers. %~\cite{ProtoBuf}. 
Specifically, metadata service and scientific discovery service 
running on each DTN %as gRPC server,
are implemented based on the client-server model using gRPC. %~\cite{gRPC}.
%\cb{Taeuk : Let's delete "as gRPC server" because that is mentioned right after comma}
%\cred{\sout{, a high performance, open-source, multi-platform, language neutral RPC framework for developing distributed applications and services}}~\cite{gRPC}. 
%Metadata service and scientific discovery service
%\cred{Why both? for what and what?}
%runs as a gRPC server on each DTN. 
The gRPC client can connect and interact with the metadata server.
%\cred{Kim: the server is running on the DTN?}. 
In our implementation, the metadata client is integrated in collaboration workspace. 
SQLite %~\cite{SQLite} 
is used as backend storage for each database shard. \tbstore{} source code consists of more than 3000 lines. 
 %
%\tbstore{} uses four metadata managers in Roud-robin policy to fairly balance the load. 
%\%cred{Kim: Why are four metadata managers? It must because your test-bed has four DTNs. But, here, you have not introduced yet your test-bed. So, here you don't have to specify the number of metadata managers.}
%Scientific discovery shard is also implemented via gRPC and co-located with metadata service.  
%We used SQLite~\cite{SQLite} as backend storage for collaboration and scientific discovery metadata.

%\cred{Kim: The above two sentences need revision. Shouldn't you write that you use SQLite for each DB shard and implement for communication from each client to the DB shard using gRPC?}

%\cred{This paragraph is good. All of the implementation-related specific content and software component in use in the previous section should be written here. }

\subsection{Experimental Setup}
\vspace{-0.05in}
\subsubsection{{Testbed}}
%In this section, we describe the testbed configurations and experimental setup as listed in Table~\ref{tab:testbed}. 
We build a testbed for scientific collaboration on top of two geo-distributed data centers equipped with Lustre~\cite{lustre} connected via high-speed Infiniband EDR (100Gbps) network. 
Table~\ref{tab:testbed} shows detail description of the testbed setup.
%and experimental setup.
%Due to limited testbed setup, each Lustre uses only single MDS and OSS server with configurations shown in Table~\ref{tab:testbed}. 
We use 2 DTNs for each data center as Lustre clients %to evaluate the performance of \tbstore{}. We 
and mount the DTNs via Linux NFS v4.0 
%~\cite{linuxNFS} 
on to the collaborator machine as shown in Figure~\ref{fig:namespace}. 
%\cred{Kim: 4 DTNs on each data center? or Total 4 dNTs? Pls clarify it.}
%The system specifications of DTNs and client node are listed in Table~\ref{tab:testbed}. 
%We mount NFS server on top of Lustre and \tbstore{} on top of NFS client.
%The testbed network connectivity is Infiniband EDR (100Gbps). 
Note our target environment is that, %\cb{(Taeuk : -> Note that in our target environment, )} 
data centers in collaboration are connected via a %\cb{(Taeuk : Let's delete this "a")} 
high-speed network such as ESNet\textquotesingle s 1Tbps network~\cite{esnet}.
We believe our testbed configuration fairly emulates this situation. 
Particularly, in such a Terabits network environment, 
the network bandwidth between the data centers is higher than the PFS bandwidth of each data center.  
To accurately emulate this situation, we have configured the Lustre bandwidth of our testbed to be smaller than the IB EDR bandwidth, as in~\cite{Kim:2015:LOD:2750482.2750488}.

%1Tbps network~\cite{Kim:2015:LOD:2750482.2750488, esnet}, 

%\cred{\sout{In particular, the parallel file system (PFS) bandwidth in each data center is lower than the network bandwidth that connects the data centers.}}

%We claim that, it is a small testbed but shares high similarity to real collaboration scenarios and same is depicted in LADS~\cite{Kim:2015:LOD:2750482.2750488}. 

% In order to fairly evaluate proposed workspace, we ensured that storage server bandwidth is not overprovisioned with respect to network bandwidth between those two data centers. 
%\cred{Duplicate contents! It has already been mentioned that the network is very fast like terabits network and storage is slower than the network. }
We compare the proposed \tbstore{} against 
%the UnionFS~\cite{unionfs}, a unification file system designed to merge several directories and file system 
a simple unification file system approach such as UnionFS~\cite{unionfs}, designed to merge several directories and file system 
branches. % based on assigned priority. 
%\cred{Kim: I don't get what is priority here.}
We implemented the prototype idea of UnionFS using FUSE for comparison with \tbstore{} and \tbstore{}-LW. In experiments, \tbstore{} refers to the use of collaboration workspace to read and write whereas, \tbstore{}-LW refers to use of the local file system namespace and can benefit with native-access support. In the rest of the paper, we refer the approach of the UnionFS as the baseline. All the experimental results show the average of multiple runs.  
We drop %\cb{(Taeuk : "page")} 
cache after each iteration of experiment from NFS mount points, DTNs, and Lustre OSSs to have authentic performance values. % picture \cb{(Taeuk : "clear picture"->"authentic result")} of performance.

\begin{figure}[!t]
	\begin{tabular}{@{}c@{}c@{}c@{}c@{}c@{}}
		\includegraphics[width=0.25\textwidth]{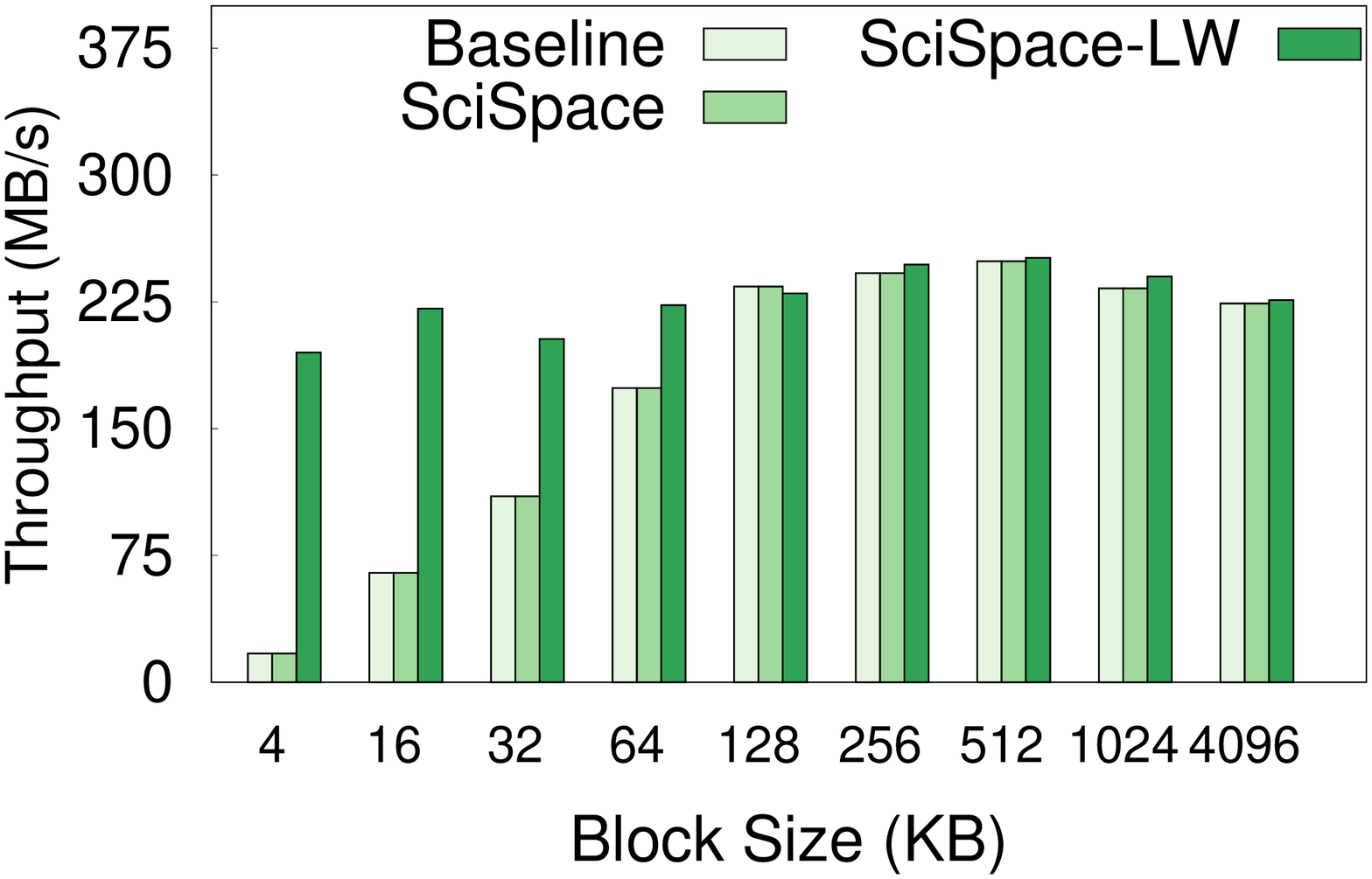} & 
		\includegraphics[width=0.25\textwidth]{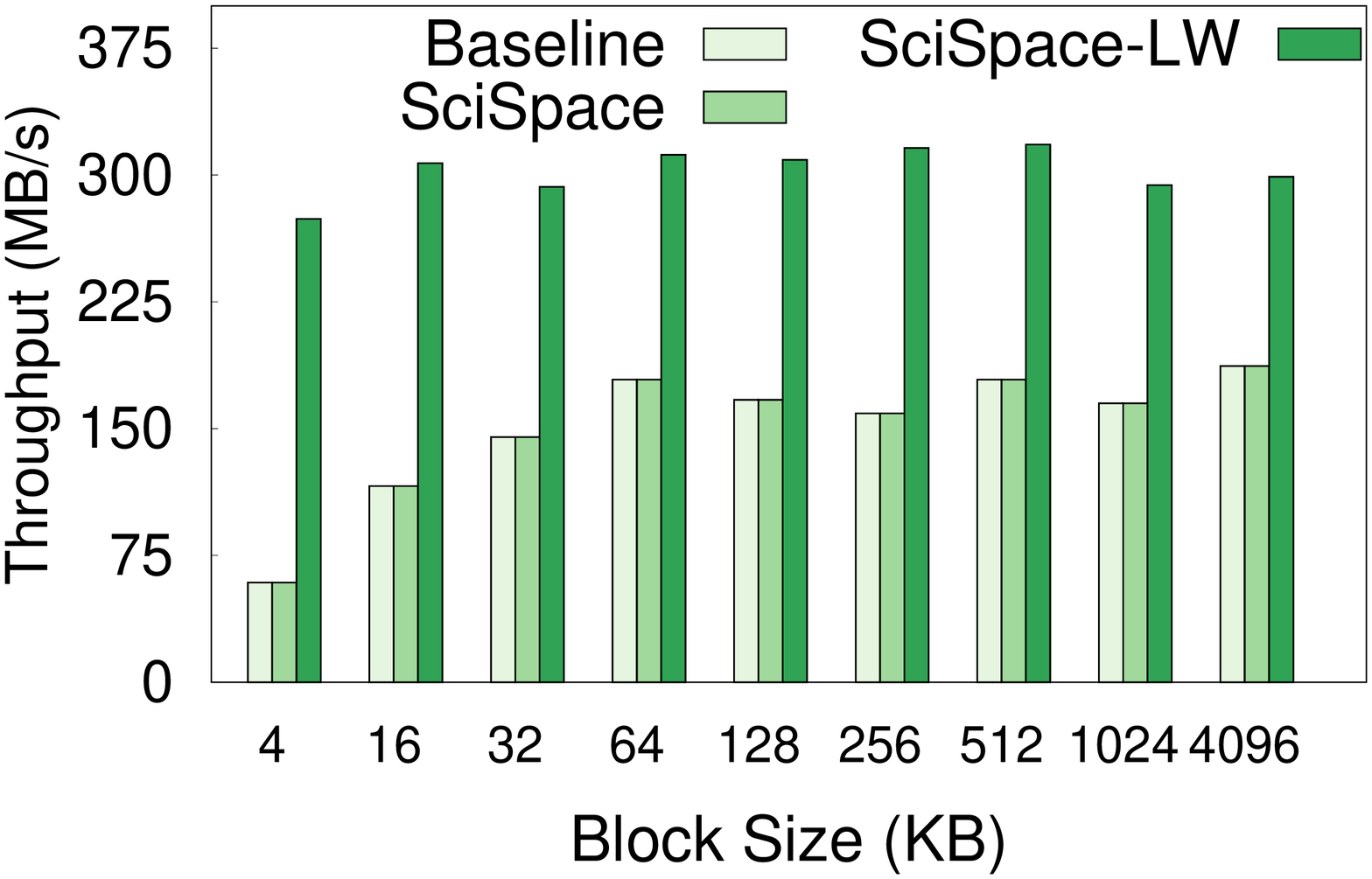} \\ 
		\small (a)  Write & 	 
		\small (b)  Read & \\	
	\end{tabular}
\vspace{-0.09in}
	\caption{\small 
	Performance analysis of \tbstore{} by varying Block-Size. }
	\vspace{-0.1in}
	\label{fig:scifs}
\end{figure}

\begin{figure}[!t]
	\begin{tabular}{@{}c@{}c@{}c@{}c@{}c@{}}
		\includegraphics[width=0.25\textwidth]{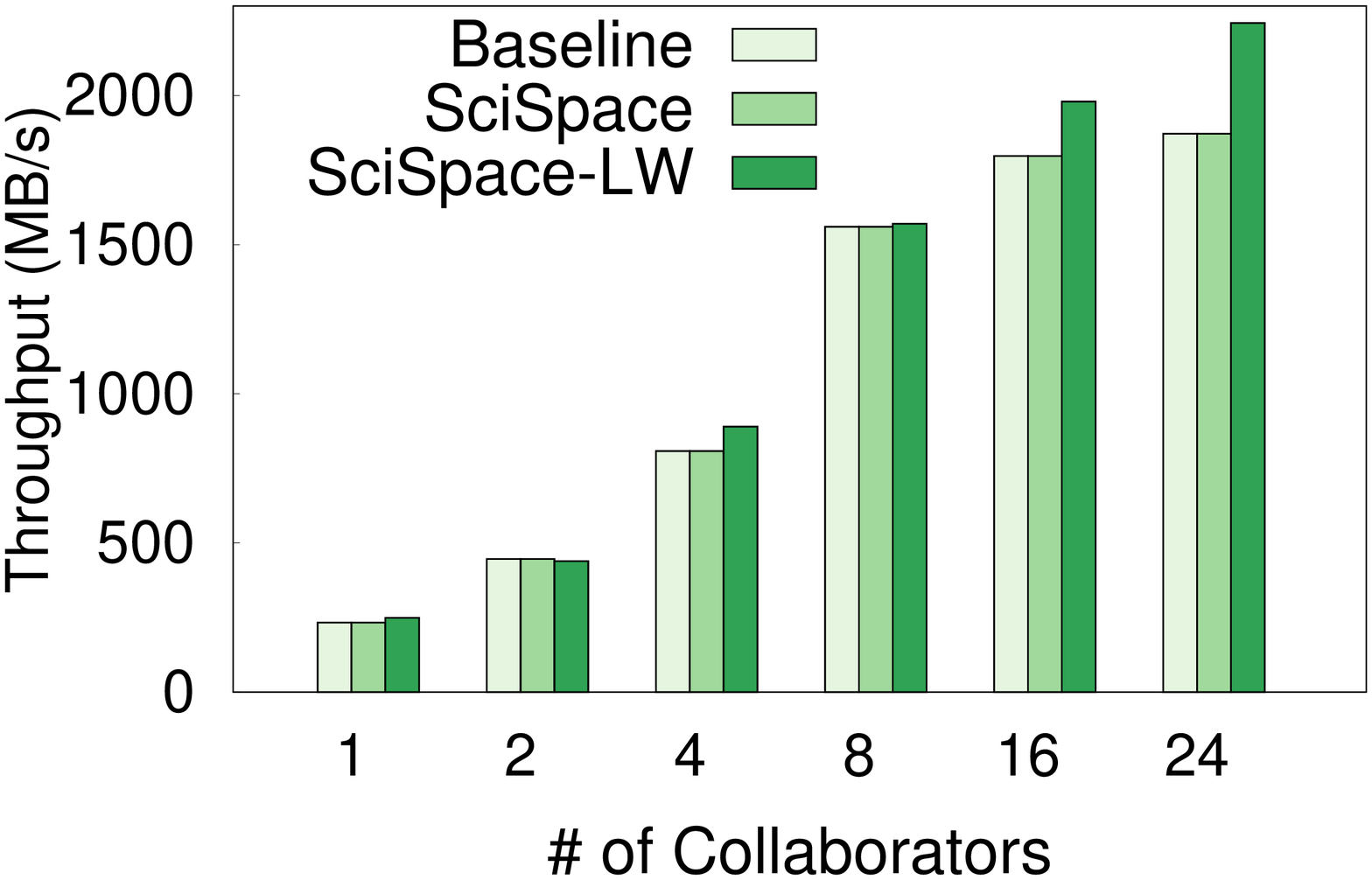}  &
		\includegraphics[width=0.25\textwidth]{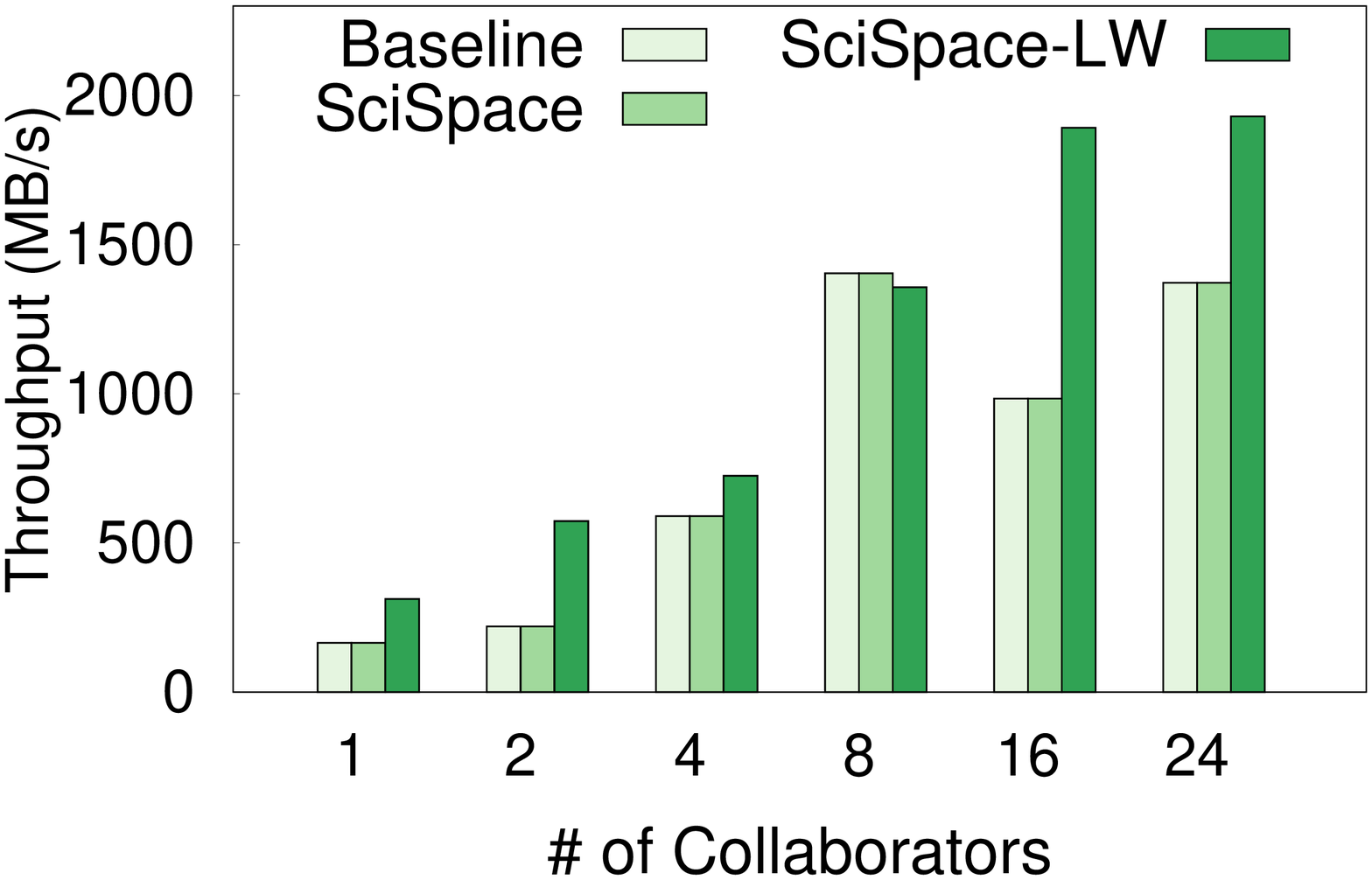}  \\	
		\small (a)  Write & 	 
		\small (b)  Read & \\
	\end{tabular}
\vspace{-0.09in}
	\caption{\small
		Performance analysis of \tbstore{} varying Collaborators. }
	\vspace{-0.25in}
	\label{fig:client}
\end{figure}

\subsubsection{Workload}
To evaluate the \tbstore{} performance, we used IOR~\cite{IOR} benchmark. We use 375 GB of synthetic dataset using IOR. The reason to use big dataset is to wipeout the caching effect.  %We used a combination of block-size, file-per-proc, direct\_io, read, and write parameters in the evaluation. 
%We adopt OpenMPI v4.0 to emulate 
%\cred{\sout{multiple collaborators}}
%the test-case of multiple collaborators to show the scalability of \tbstore{}. 
%\cred{Kim: this sentence is broken. You mean you used the version of OpenMPI for IOR or you used two different benchmarks, IOR and OpenMPI? OpenMPI is not a benchmark program.}
%We configure IOR to use 375 GB synthetic workload to evaluate \tbstore{} file system namespace. 
For real collaboration activities, we use scientific HDF5 datasets comprised of the ocean surface data measured at different time period across geo-distributed locations by different scientific instruments. 
We downloaded the dataset of size 116 GB (4600 files) from MODIS-Aqua~\cite{modis}%(Moderate Resolution Imaging Spectroradiometer)
. MODIS plays a vital role to predict global changes accurately enough to assist policymakers in making decisions concerning the protection of our climate~\cite{modis}.
%\cb{Taeuk : There's no mention about the source of other 259GB data} 
\begin{comment}
we used different workloads based on HDF5~\cite{hdf5}. \cb{Add ocean and temperature stuff etc} HDF5 is a completely new Hierarchical Data Format specially designed for scientific and research communities due to unstructured data which naturally does not fit into relational tables. The single HDF5 file can have multiple different objects stored in hierarchy~\cite{hdf5}. %We categorized the workload into two major categories, i.e., the small workload with an average file size less than 2MB and big workload with average file size greater than 200MB. 
We obtained the real HDF5 dataset from of size 116 GB  with total number of 4600 files. The size-based distribution of HDF5 files is shown in Figure~\ref{size-distribution}.
\cred{Kim: did you use these tools to show results? Write more about the collaboration activities. Readers want to understand what you have done by just reading this sentence to see what will come next.}
\end{comment}
%Next, we list the 
We use two HDF5 applications, i.e., H5Diff and H5Dump in order to emulate real collaboration activities. 
%\cred{Kim: This sentence is broken.}
% are commonly used for HDF5 datasets. 
%collaboration activities for scientific datasets.
%\cred{Kim: This sentence does not make sense. It does not explain why you used H5Diff and H5Dump. Didn't you want to say you used them to evaluate the active operation on each DTN? If so, make sure you explain what the active operation is in the pervious sections.}

\begin{comment}
\squishlist
\item
\textit{h5Diff}: 
\item
\textit{h5Dump}: 
\item
\textit{h5Copy};
\item
\textit{h5import}:
\item
\textit{h5stat}: 
\squishend
\end{comment}

%\vspace{-0.1in}

\subsection{Scientific Collaboration Workspace}
\vspace{-0.05in}

\begin{comment}
\cb{Khan: experiment \# 1 read, write with varying block size in Lustre, NFS atop of Lustre and SciFs}
\cb{Khan: experiment \# 2 read, write varying client size in Lustre, NFS atop of Lustre and SciFs}
\cb{Khan: experiment \# 3 applications runtime in Lustre, NFS atop of Lustre and SciFs}
\cb{Khan: experiment \# 4 mdtest for files and directories}
\cb{Khan: experiment \# 5 export utilitiy experiment}
\cb{Khan: experiment \# 6 load balancing of Scifs for different policies}
\end{comment}

%\cred{\sout{To evaluate the \tbstore{} framework,}}
To evaluate the performance overhead of \tbstore{} framework,
we run two sets of experiments (read, write) and compare baseline with two variants of the proposed framework, i.e., \tbstore{} and \tbstore{}-LW. (quoted as native-access). %\tbstore{} uses collaboration workspace to read and write whereas, \tbstore{}-LW uses the local file system namespace and can benefit with local-writes support.
%\cred{Kim: as commented earlier, this paragraph can be moved early in Experimental Setup?}
%\cred{Kim: Awais, did you explain what \tbstore{}-LW (quoted as Local-Write) is before? If not, here explain what \tbstore{} and the other, \tbstore{}-LW are. }

%\cred{Kim: this can be moved to workload section too.}
%Figure~\ref{fig:scifs} (a)(b) show the experimental results with varying block-size with a single client. 

%to evaluate the performance overhead. We compare \tbstore{} with baseline.  %(Linux's network file system NFS v4.0). 
%\cred{ Let's call the baseline that I/O and indexing are performed through SciSpace as described earlier. Otherwise, your evaluation results show no more than just analyzing the FUSE overhead.}
%By default, baseline uses both client and server-side cache. However, we disable the client-cache when running IOR benchmark.  

In Figure~\ref{fig:scifs} (a)(b), %\cb{Taeuk : In Fig 7, between 1024 and 4096 on x-axis, there is no empty space, so it can be read as 10244096} 
we investigate the impact of block size in both write and read operations with a single collaborator. We observe that when the block size is less than 16KB, the write and read performance degrades in both baseline and \tbstore{} as compared to \tbstore{}-LW. The reason for the decrease in read and write operations is due to small-size transfer requests, FUSE layer overhead, and metadata contact points. Whereas, \tbstore{}-LW shows higher performance due to local-writes support and low metadata contact points.
%\cred{Kim: You missed to explain why \tbstore{}-LW does not degrade.}
However, as we increase the block size, the write and read performance increases in all three approaches. %\cb{Taeuk : I think SciSpace-LW do not show clear increment of throughput in Fig7}
In Figure~\ref{fig:scifs}, the maximum throughput %\cb{s} 
achieved by the baseline and \tbstore{} is %\cb{(Taeuk : "is" -> "are")} 
same at a block-size of 512KB however, the \tbstore{}-LW shows higher performance in all test cases, in particular ranging from small block-size 4KB up to 512KB. The performance improvement window lies in range from  2\% up to 70\% when moving from big block-size to smaller block-size. %\cb{Taeuk : Isn't it "from smaller block-size to big block-size"?} 
The average performance improvement of all write test-cases is 16\%. However, for read test-case, \tbstore{}-LW shows a consistent performance improvement in all test cases with an average of 41\%. The performance degrades in baseline and \tbstore{} due to several factors, first additional metadata querying for stat, second  FUSE invokes five operations serially, getattr, lookup, create, write and flush % \cb{(Taeuk : Are "create", "write" and "flush" FUSE function called in read workload case?)} 
and third, user and kernel space context switching overhead cannot be ignored. Whereas, in \tbstore{}-LW case, we allow collaborators to write to local file system namespace and push the unsynchronized metadata to \tbstore{}. We have no additional metadata querying and no FUSE overhead in SciSpace-LW.

Next, we perform the experiment to show scalability of \tbstore{} collaboration workspace by increasing number of collaborators. 
Figure~\ref{fig:client} (a)(b) shows the impact of multiple collaborators in both read and write operation of all three approaches, i.e., baseline, \tbstore{} and \tbstore{}-LW. 
We use the same dataset of size 375GB via IOR and fix the block-size to 512KB to benefit the baseline and \tbstore{} approach as compared to \tbstore{}-LW. The results stand different from our observations in the previous experiment Figure~\ref{fig:scifs}. As we vary the number of collaborators, the baseline, \tbstore{}, and \tbstore{}-LW show %\cb{(Taeuk : Let's delete this "s")} 
a consistent performance improvement. The reason for this improvement is manifold. First, baseline and \tbstore{} get the benefit of NFS caching at server and Lustre OSS cache and parallelism. Second, due to %\cb{(Taeuk : Let's replace from "is" to "due to")} 
effective and load-balanced utilization of available DTNs, i.e., in the baseline, we allocate each DTN equal priority and in \tbstore{}, we configure round-robin request placement policy. However, \tbstore{}-LW, we divide the number of collaborators on each DTN. Whereas, our \tbstore{}-LW cannot benefit with NFS caching because it directly runs on local data center namespace and can only utilize the parallelism of deployed Lustre at the data center. The maximum performance boost when 24 collaborators are active in collaboration is; for write test-case, 16\% and read test-case shows 28\% boost when compared to baseline and \tbstore{}. However, we consider it is important to show the reason for read performance degradation when collaborators number varies from 8-16. The reason behind is NFS caching. So, in baseline and \tbstore{} when the cache is full, the flush operation is invoked and all the write I/Os get slow due to multi-level cache (NFS cache, Lustre OSS) flush operation in progress. On the contrary, \tbstore{}-LW requires only single cache flush (Lustre OSS).

\begin{figure*}[!th]
	\begin{tabular}{@{}c@{}c@{}c@{}c@{}c@{}}
		\includegraphics[width=0.34\textwidth]{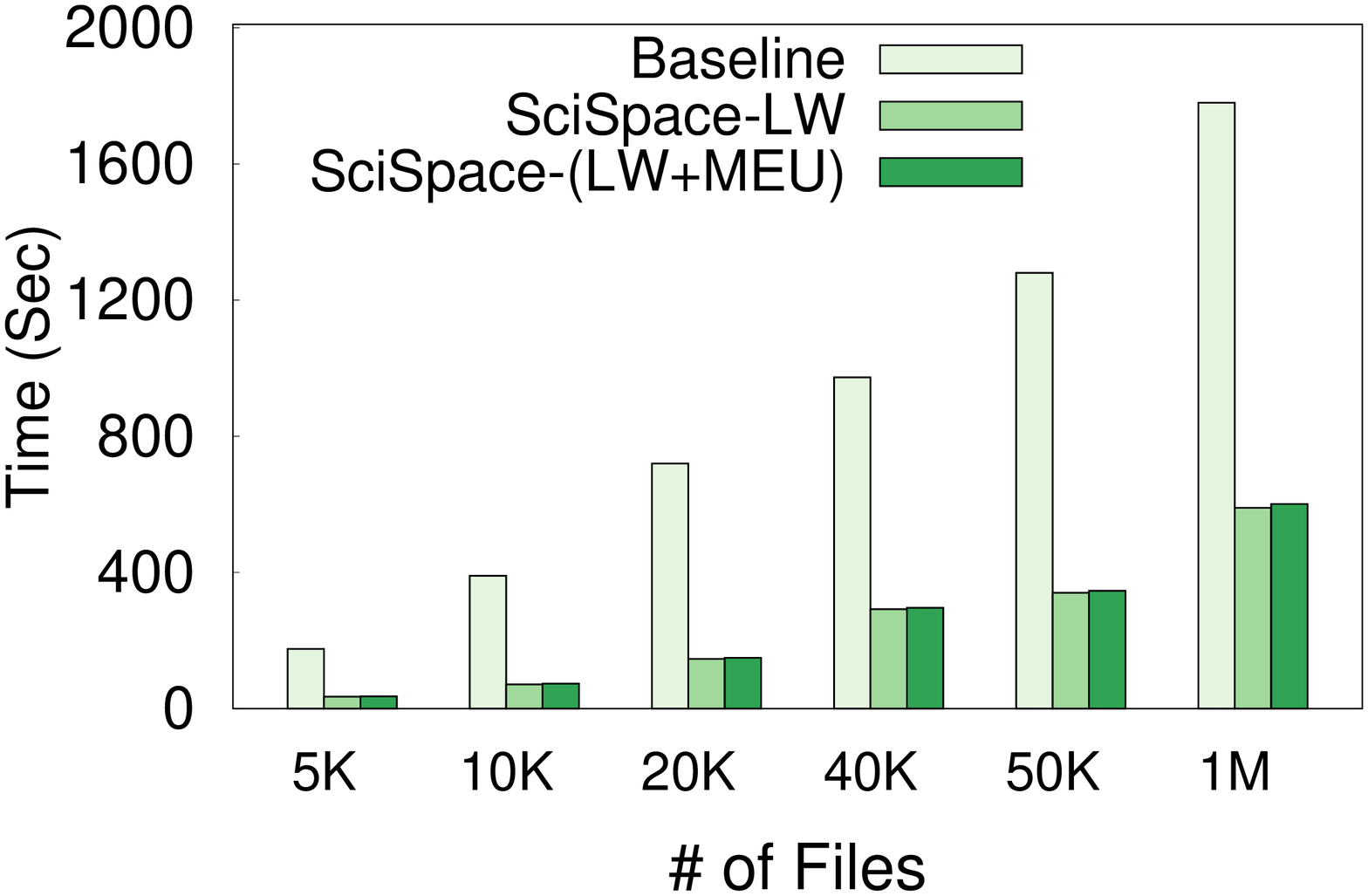} & 
		\includegraphics[width=0.34\textwidth]{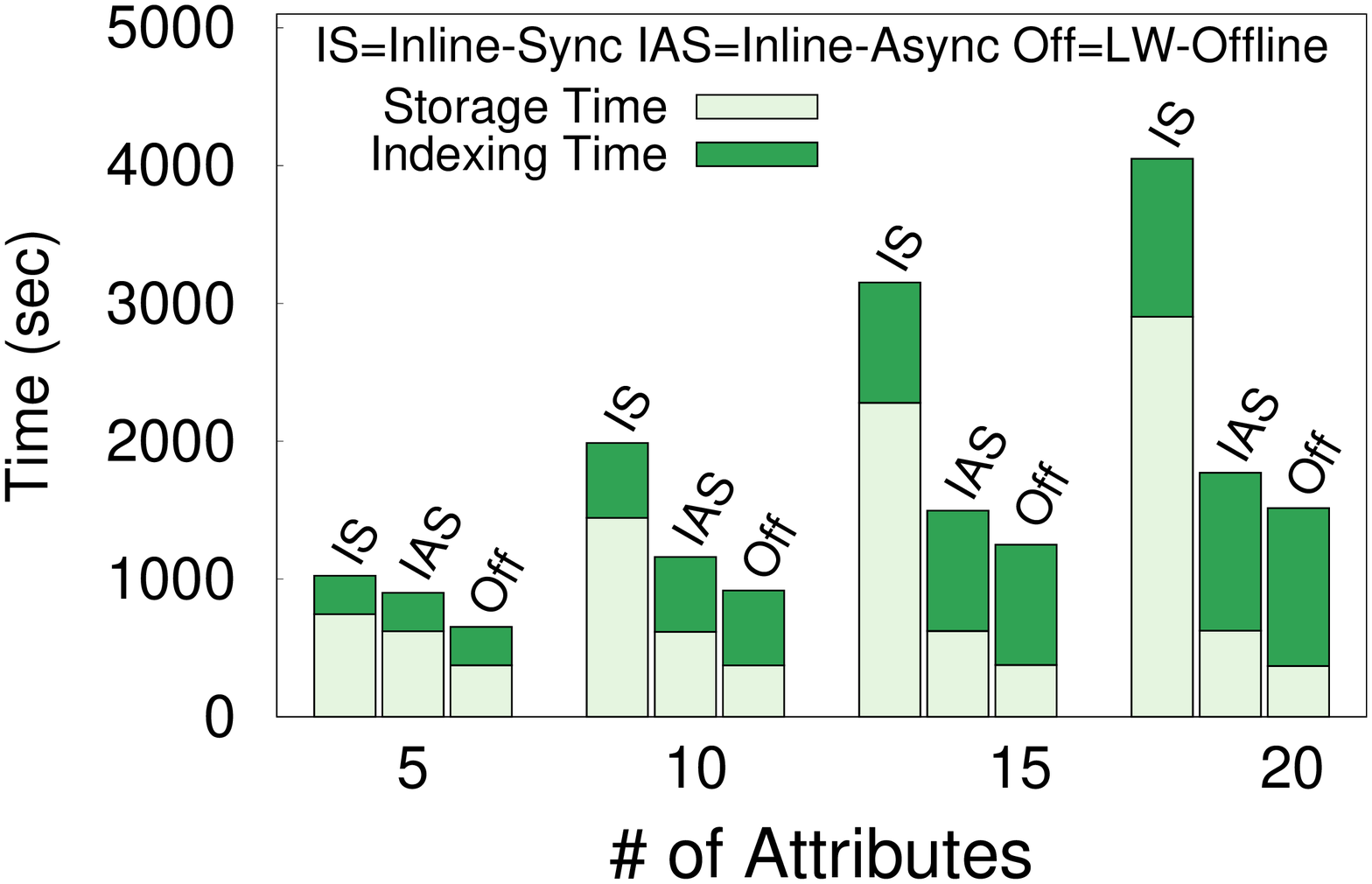} &
		\includegraphics[width=0.34\textwidth]{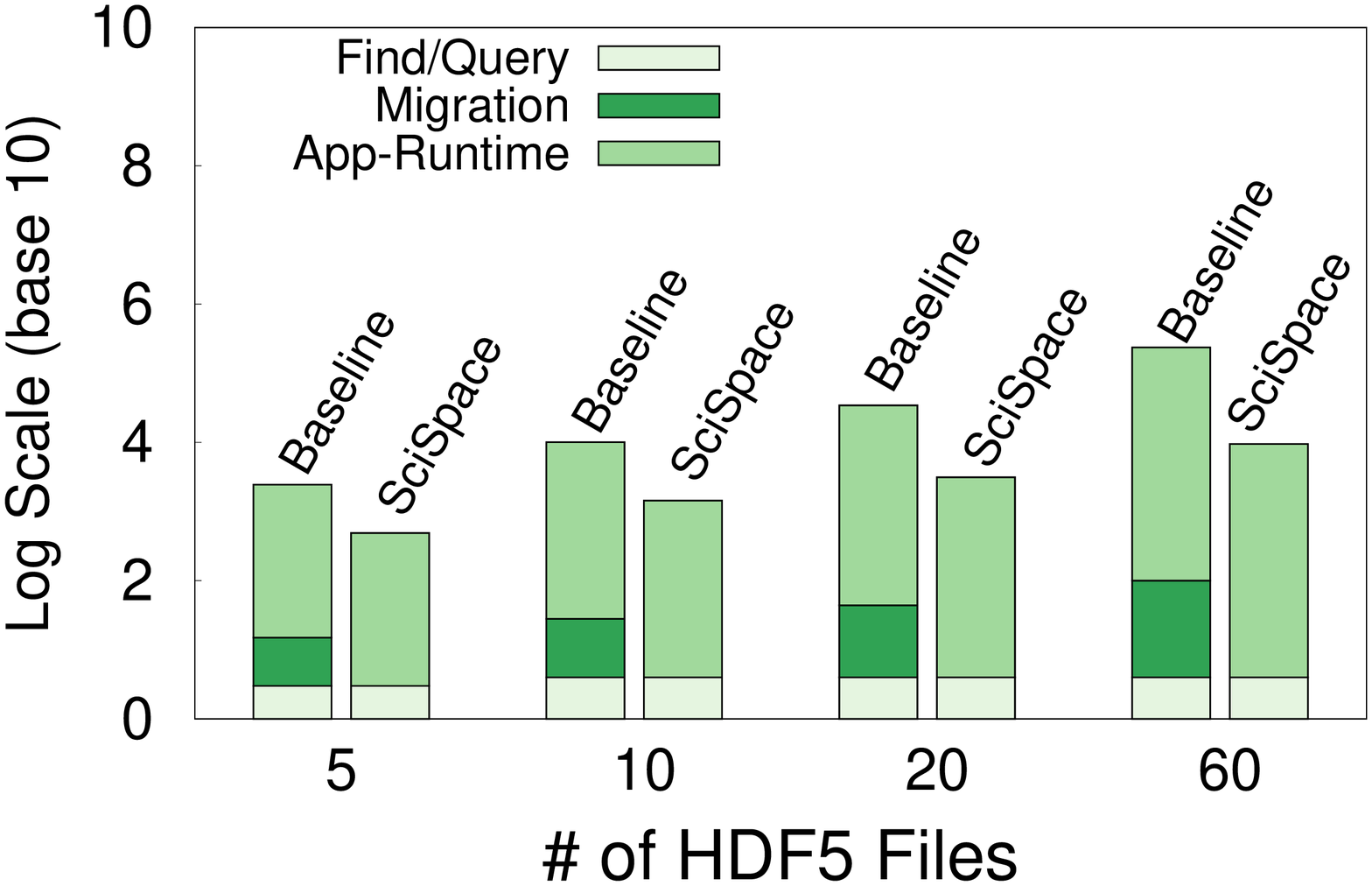} \\ 
		\small (a) MEU Analysis &
		\small (b) Indexing Modes &
		\small (c) HDF5 Tools\\
	\end{tabular}
	\vspace{-0.05in}
	\caption{\small 
		\tbstore{}: Metadata Export Utility, Indexing Modes and End-to-End Collaboration Performance with real HDF5 tools.}
	\vspace{-0.1in}
	\label{fig:fsscan}
	\label{fig:tagging}	
	\label{fig:hdf5}
\end{figure*}

\begin{comment}
whereas,  baseline runs all clients running on a single host where it is paired with a DTN at remote, whereas in SciFS, it can fully utilize all DTNs at remote as it is a 1-to-many mapping. 
}
\end{comment}
%On the contrary, \tbstore{} can utilize multiple DTNs (unification) and balance jobs in the round-robin fashion. 

%We configured the load-balancing policy as round-robin for all experiments in this section. 
%\cred{Here you go. How the load-balancing is implemented with SciSpace should have been explained in the preceding section!}
%\tbstore{} distributes the data to each of four data transfer nodes and benefits with server-side caching, as no cache contention occurs as compared to baseline. Thus, we claim that \tbstore{} is highly feasible and practical solution to be deployed in real collaborations.

\subsection{Metadata Export Utility}
\vspace{-0.05in}

\begin{comment}
\cred{Kim: shouldn't this result come before showing the results of Attribute-based Data Discovery? Swap them to show.}
\cred{Kim: Fig6 results do not seem to be aligned with those in Fi4. In Fig4, results show SciFS overhead is not shown that bad compared to native baseline approach. I know there is subtle difference in evaluation methodology. Pls check two results in comparison and re-run experiments if necessary, or add more writing in detail.}
\end{comment}

Metadata Export Utility (MEU) performance relies on the number of files,  irrespective of file size. Our realistic dataset contains 4600 files (116 GB), which we believe is not sufficient to clearly show the performance of MEU. 
To show the effectiveness of the proposed approach using a single collaborator, we define a simple workflow. We create a zero-size file (count 5K-1M) via baseline, \tbstore{}-LW and execute the MEU on top of \tbstore{}-LW (Figure~\ref{fig:fsscan} (a)) to synchronize the metadata of files such as filename and location (File Mapping Schema in Figure~\ref{fig:shard}). The baseline approach uses the common FUSE-based collaboration workspace. In \tbstore{}-LW, all the files are created via local file system namespace however, it does not include the MEU export overhead. Whereas, \tbstore{}-(LW+MEU) includes the use of local file system namespae and MEU export overhead as well.
%\cred{Kim: Explain the difference between SciSpace-LW and SciSpace-(LW+MEU). A baseline SciSpace allows to go through a FUSE common namespace for all I/O operations? }
The experimental results are shown in Figure~\ref{fig:fsscan} (a). We observe that baseline creates a huge overhead which comes from increased contact points between collaboration workspace and metadata service. Each of the file system calls (such as attr, access, create, open) requires assistance from metadata service. Whereas, \tbstore{}-LW requires no such additional metadata assistance. However, MEU recursively iterates the directories and create the a list of unsynced files and send message to metadata service on DTN. The \tbstore{}-LW and \tbstore{}-(LW+MEU)% \cb{(Taeuk : -> SciSpace-LW and SciSpace-(LW+MEU))} 
show %s\cb{(Taeuk : Let's delete "s")} 
a linear performance pattern. In MEU, we batch all the requests and send single RPC call to metadata service to minimize the message packing overhead.

%\cred{Awais, NativeAccess+MDExport, how about proposing an idea of NativeAccess + MDExport? In other words, our story is we understand it is important to create a collaboration workspace for scientists' collaboration promotion, however as writing directly to SciSpace has high performance overhead such as FUSE, so we propose a new idea that uses NativeAcces and MDExport in combination. Because Figure8 shows quite positive results with NativeAccess+MDExport. There is no reason to use the SciSpace for I/O due to high performance overhead. }

\iffalse
\begin{figure}[!t]
	\centering
	\includegraphics[width=0.40\textwidth]{./gnuplot/expr3.eps}  
	\caption{ \small
		Performance analysis of metadata export utility count increasing file count.
		%\cred{Kim: Change SciFS to SciSpace in labels}
	}
	\vspace{-0.15in}
	\label{fig:fsscan}
\end{figure}
\fi

%\vspace{-0.05in}

\subsection{Scientific Discovery Service}
\vspace{-0.05in}
%\cred{From Fig9, we see NativeAccess-Interleaving mode shows the best performance. Even though I don't like the name you call NativeAccess-Interleaving, it follows the idea as mentioned in the proceeding results, using the NativeAccess+MDExport. Again, how about making this an idea?}

In this section, we show the performance of multiple metadata extraction modes. For this experiment, we use the 4 collaborators and real scientific HDF5 datasets (116 GB). We extract all the attributes (Search Attribute in Table~\ref{tab:query}) from HDF5 files
%\cred{Kim: What are these attributes? Are the attributes in the HDF5 file format? }
along with file system metadata (pathname, size, time, inode number etc.). We specifically present {Inline-Sync}, {Inline-Async}, and {LW-Offline}. We described each mode in detail in Section~\ref{sec:discovery}. The Inline-Sync and Inline-Async use \tbstore{} collaboration workspace, whereas LW-Offline uses the local data center namespace. It indexes the files and update SDS shard accordingly.

Figure~\ref{fig:tagging} (b) shows the time breakdown analysis of all the data discovery modes. As expected, the Inline-Async and LW-Offline perform better with an improvement factor of 12\% and 36\% with 5 attributes when compared to Inline-Sync. Whereas, when 20 attributes are used, the performance boosted up to 56\% in Inline-Async and 62\% LW-Offline. The high time taken by Inline-Sync is mainly derived from I/O blocking. A single write I/O waits until all the indexing operations are complete. The indexing operations include opening HDF5 file, extracting metadata attributes, and recording the attributes in the database. Also, when we compare Inline-Async and LW-Offline, the performance in earlier one is 56\% and later one is 62\% as compared to Inline-Sync when 20 attributes are used. The reason for negligible performance overhead in Inline-Async as compared to LW-Offline is the result of additional gRPC calls and protobuf messages for enqueuing the index messages. However, LW-Offline operates directly on the local file system namespace and incurs no added messaging overhead.

%\begin{comment}
%Here, we clearly observe a constant pattern followed by each data discovery mode. Inline-sync mode shows high performance overhead as compared to inline-async and offline-LN. The reason for performance degradation includes; First, Inline-Sync mode uses the FUSE layer. Second, when FUSE write operation finishes the actual data write, a request is sent to indexing server to index the file. The index server extracts the metadata attributes defined in HDF5 file and push attributes in DB. When index step is complete, the index server sends ack message to the sender. The key observation here is; the request waits until the indexing operation is finished. but only messages are passed and no notable overhead is observed. The Offline-LN utilize the local namspace path and incurs no additional FUSE overhead, whereas the file indexing is done in offline mode.
%\end{comment}

\iffalse
\begin{figure}[!t]
	\centering
		\includegraphics[width=0.40\textwidth]{./gnuplot/expr5.eps}  \\
	\caption{ \small
	Performance analysis of multiple indexing modes. The x-axis denotes \# of attributes.
	}
	\vspace{-0.1in}
	\label{fig:tagging}	\label{fig:hdf5}
\end{figure}
\fi

Next, we discuss the search query latency. We measure the search query latency using 4 collaborators, each produces four types of 1000 queries. We select each query based on the defined attributes in the real HDF5 dataset, i) search the files generated at a certain location, ii) search the files with the particular instrument, iii) search the files including specific date, iv) search the files generated in day or night. We populate the SDS shards with indexes and show latency by varying hit-ratio. The hit-ratio is defined as the number of matching tuples in SDS shard over the total number of tuples in shard. The average latency of each query is listed in Table~\ref{tab:query}. We have seen that when hit-ratio is less, i.e., the number of matching entries are only 25\% of total entries, the query latency is very short up to 8-9 seconds. However, when we vary the hit-ratio to 100\%, the high latency is experienced in all search queries. This increase in query latency is the result of message packing and unpacking at SDS. The SDS translates the request message into SQL query and finds the required attributes in SDS shard, then query results are packed in a message and sent over the network. When the number of records returned in the SQL query is high, then latency increases. This internal message packing overhead leads us to show the hit-ratio comparison.

\begin{table}[!t]
	\small
	\caption{\small Search query latency (in seconds) by varying Hit-Ratio.}
	\vspace{-0.05in}
	\begin{tabular}{|ll|l|l|l|l|l|l|l|l|l|l|l|l|l|}
		\hline
		\multirow{2}{*}{\textbf{Search Attribute}} & 
		\multirow{2}{*}{\textbf{}} & 
		\multicolumn{1}{c}{} &
		\multicolumn{2}{c}{\textbf{Hit-Ratio}} &
		\multicolumn{2}{c|}{} \\ \cline{3-7} 
		& & 0\% & 25\% & 50\% & 75\% & 100\% \\ 
		\hline
		Location  (Text)& & 3.6 & 9.7 & 14.6 & 19.5  & 24.5  \\
		\hline
		Instrument (Text) & & 	3.8 & 9.5 & 14.7 & 19.7 & 24.5  \\
		\hline
		Date (Text) && 3.9 & 9.6 & 14.8 & 19.7 & 24.6 \\
		\hline
		Day or Night (Int) & & 3.2 & 8.9 & 14.1 & 18.9 & 23.9 \\
		\hline
	\end{tabular}
	\vspace{-0.2in}
	\label{tab:query}
\end{table}

\iffalse
\begin{figure}[!t]
	\centering
	\includegraphics[width=0.35\textwidth]{./gnuplot/expr7.eps}  
	\caption{ \small
		Performance of \tbstore{} with real HDF5 applications. The x-axis denotes number of files.
		\cred{Kim: change the label name of SciFS to SciSpace}
	}
	\vspace{-0.1in}
	\label{fig:hdf5}
\end{figure}
\fi

\subsection{End-to-End Analysis for Scientific Collaborations}
\vspace{-0.05in}
We conduct the experiments to compare end-to-end analysis times 
between baseline approach and \tbstore{} with real HDF5 tools
%We measure an end-to-end analysis time with real HDF5 tools 
%Next, we conduct the experiment with real HDF5 tools 
such as H5Diff (computing the difference between two HDF5 files) and H5Dump (converting HDF5 file to ASCII file). 
%H5Diff and H5Dump are applications to compute the difference between two files and to convert HDF5 file to ASCII file respectively. 
%We compare \tbstore{} with baseline approach which requires 
In the baseline approach, it 
first finds the datasets on different data centers, then migrates the datasets from all locations to local data center and run applications. 
In particular, the search time increases as the number of files searched increases because it only allows file-name based search. 
On the other hand,
% however, 
collaboration namespace gives benefit in terms of first two steps;
%, i.e.,
first, query time is constant irrespective of data size and file count. 
Second, no-migration is required because application can run directly on searched dataset without transferring datasets to the local data center. 
%Moreover, the query time in the baseline denotes the Linux \textit{Find} command, which increases the search time when a number of file increases. 
%The evaluation of two HDF5 applications, H5Diff, and H5Dump is shown in Figure~\ref{fig:hdf5} using logscale to clearly present the overhead of each step. 
Figure~\ref{fig:hdf5} (c) shows the result of H5Diff application. \tbstore{} shows lower end-to-end run times than baseline for all cases of different files. %Whereas, baseline bbcp uses a single pair of DTNs, whereas SciFS can fully utilize all DTN pairs.
We observe the same performance trend for the H5Dump application %\cb{(Taeuk : with H5Diff)}
, however due to page limit, we do not show the H5Dump results.
\begin{comment}
\cred{Kim: first write SciFS shows lower end-to-end run times than baseline for all cases of different files tested.
Also, mention in the baseline bbcp uses a single pair of DTNs, whereas SciFS can fully utilize al DTN pairs.
}%We observe \tbstore{} shows lower runtimes than baseline irrespective of number of files. 
\end{comment}}
We observe an exponential increase in baseline approach due to migration time. 
\cred{Kim: don't we see the same exponential increase in SciFS? If not why? If so, then, mention they show the same.}
However, \tbstore{} frees collaborator from manual data transfers and reduces the manual scientific workflow in collaborations.
\cred{Kim: then write here we had the same observational trend from H5Dump applications.}
We encounter the same pattern in both applications. 

\begin{comment}
\cred{Kim: you need more to write about the results and analysis. What is hit ratio? How can you conclude that it's scalable? Scalability can be shown only by evaluating query processing time w.r.t increased client count. Refer to Tagit paper. 
}
\end{comment}

%% file: conc.tex
\section{Conclusion}
\label{sec:conc}
\vspace{-0.05in}
%The increasing collaborations among geospatial data centers require a notion of the common workspace, where all collaborators can easily view and share data. Moreover, in case of multiple data centers, the existing remote login utilities (ssh), and data transfer tools cannot satisfy the needs of collaborators. To address these needs, we propose \tbstore{}, a scientific collaboration file system which offers a virtually unified common namespace  to collaborators in multi-data center environment. Additionally, \tbstore{} also offers the scientific discovery service which reduces the traditional scientific workflows by efficiently extracting the desired datasets via querying the file system. \tbstore{} supports the use of local/native datacenter namespace bypassing the global namespace. We build collaboration between two data centers equipped with Lustre connected via Infiniband to evaluate \tbstore{}. The evaluation supports the effectiveness of \tbstore{} in real collaborations.

%%% NEW COMMENTED
%The increasing collaborations among geospatial data centers demand a notion of the shared workspace, where all collaborators can easily view and share data. Further, in case of multiple data centers and super-abundant files, the existing tools such as ssh and transfer tools (bbcp) cannot meet the specific needs of collaborators such as efficient retrieval of desired information. 

In this work, we propose \tbstore{}, a Scientific Collaboration Workspace which offers a virtually unified common workspace to collaborators in multi-data center collaborations. \tbstore{} supports native-data access to achieve high-performance via metadata export protocol. Scientific discovery service reduces the scientific workflows by efficiently extracting the desired datasets via offering search query-like utility. We evaluated \tbstore{} on top of two small-scale geo-distributed HPC data centers connected via Infiniband and equipped with Lustre. The evaluation confirms the usefulness of the \tbstore{}. 
% and shows 36\% performance improvement when deployed in real collaborations.